\PassOptionsToPackage{colorlinks,citecolor=blue,urlcolor=blue}{hyperref}
\documentclass[11pt]{article}

\usepackage[margin=1in]{geometry}
\usepackage{amsthm,amsmath,amsfonts,amssymb}
\usepackage[authoryear]{natbib}
\usepackage{xcolor}
\usepackage{float}
\usepackage{booktabs}
\usepackage{tikz}
\usepackage{orcidlink}
\usetikzlibrary{bayesnet}
\usetikzlibrary{positioning}
\usepackage[colorlinks,citecolor=blue,urlcolor=blue]{hyperref}
\usepackage{graphicx}

\usepackage{etoolbox}
\usepackage{bbm}

\def\oper#1{\csdef{#1}{\operatorname{#1}}}
\forcsvlist\oper{GL,SL,deg,Hom,End,Vect,Mod,Rep,Ker,Im,Id}
\def\prob{\operatorname{P}}

\def\Mult{\operatorname{Multinomial}}
\def\Dir{\operatorname{Dirichlet}}
\def\Bern{\operatorname{Bernoulli}}
\def\pg{\operatorname{PG}}
\def\Ex{\operatorname{E}}

\def\diag{\operatorname{diag}}

\def\cl{\operatorname{cl}}
\def\nsub{\operatorname{n}}

\def\defbb#1{\csdef{b#1}{\mathbb{#1}}}
\forcsvlist\defbb{A,B,C,D,E,F,G,H,I,J,K,L,M,N,O,P,Q,R,S,T,U,V,W,X,Y,Z}

\def\defcal#1{\csdef{c#1}{\mathcal{#1}}}
\forcsvlist\defcal{A,B,C,D,E,F,G,H,I,J,K,L,M,N,O,P,Q,R,S,T,U,V,W,X,Y,Z}

\def\boldc{\mathbf{c}}

\def\boldm{\mathbf{m}}

\def\boldv{\mathbf{v}}
\def\boldw{\mathbf{w}}
\def\boldx{\mathbf{x}}
\def\boldy{\mathbf{y}}
\def\boldz{\mathbf{z}}

\def\boldC{\mathbf{C}}

\def\boldI{\mathbf{I}}

\def\boldS{\mathbf{S}}

\def\boldV{\mathbf{V}}

\def\boldX{\mathbf{X}}
\def\boldY{\mathbf{Y}}
\def\boldZ{\mathbf{Z}}

\def\bsal{\boldsymbol{\alpha}}
\def\bsbet{\boldsymbol{\beta}}
\def\bsgam{\boldsymbol{\gamma}}

\def\bslam{\boldsymbol{\lambda}}

\def\bsthet{\boldsymbol{\theta}}

\def\bsom{\boldsymbol{\omega}}

\def\bskap{\boldsymbol{\kappa}}

\def\bsphi{\boldsymbol{\phi}}

\def\bspi{\boldsymbol{\pi}}
\def\bsnu{\boldsymbol{\nu}}
\def\bsrho{\boldsymbol{\rho}}

\theoremstyle{plain}

\theoremstyle{definition}

\title{Bayesian Latent Class Regression and Variable Selection with Applications to Sleep Patterns Data}

\author{
  Matthew Heaney\,\orcidlink{0009-0005-4286-1559}\thanks{School of Computer Science and Statistics, Trinity College Dublin. Email: \href{mailto:heaneym@tcd.ie}{heaneym@tcd.ie}} \and
  Olive Healy\,\orcidlink{0000-0002-2119-4129}\thanks{School of Psychology, Trinity College Dublin.} \and
  Jason Wyse\,\orcidlink{0000-0003-1391-7371}\footnotemark[1] \and
  Arthur White\,\orcidlink{0000-0002-7268-5163}\footnotemark[1]
}

\date{} % or specify a date

\begin{document}

\maketitle

\begin{abstract}
Sleep difficulties in children are heterogeneous in presentation, yet conventional assessment tools like the Children's Sleep Habits Questionnaire (CSHQ) reduce this complexity to a single cumulative  score, obscuring distinct patterns of sleep disturbance that require different interventions. Latent Class Regression (LCR) models offer a principled approach to identify subgroups with shared sleep behaviour profiles whilst incorporating predictors of group membership, but Bayesian inference for these models has been hindered by computational challenges and the absence of variable selection methods. We propose a fully Bayesian framework for LCR that uses Pólya-Gamma data augmentation, enabling efficient sampling of regression coefficients. We extend this framework to include variable selection for both predictors and item responses: predictor variable selection via latent inclusion indicators and item selection through a partially collapsed approach. Through simulation studies, we show that the proposed methods yield accurate parameter estimates, resolve identifiability issues arising in full models, and successfully identify informative predictors and items while excluding noise variables. Applying this methodology to CSHQ data from 148 children reveals distinct latent subgroups with different sleep behaviour profiles, anxious nighttime sleepers, short/light sleepers, and those with more pervasive sleep problems, with each carrying distinct implications for intervention. Results also highlight the predictive role of Autism Spectrum Disorder diagnosis in subgroup membership. These findings demonstrate the limitations of conventional CSHQ scoring and illustrate the benefits of a probabilistic subgroup-based approach as an alternative for understanding paediatric sleep difficulties.
\end{abstract}


\section{Background}

\subsection{Introduction}
Sleep health is recognised as crucial in determining overall health \citep{meaklim_sleep_2020} and the impact sleep insufficiency has on the human body is well understood. Sleep plays an important role in physical and mental wellbeing in children, impacting the immune system, capacity for learning and overall mood (\citet{antczak_physical_2020}; \citet{matricciani_childrens_2019}; \citet{perry_raising_2013}). This has lead to public health experts recommending that sleep health is given the same attention as nutrition and physical activity in public health guidelines \citep{chaput_sleeping_2018}.  
Paediatric sleep difficulties are common, with up to 40\% of neurotypical children experiencing sleep problems at some point up to adolescence \citep{meltzer_sleep_2014}, and neurodiverse children having a higher propensity to experience difficulties. It is well documented that the presentation of sleep difficulties can be heterogeneous in nature. 

The Children's Sleep Habits Questionnaire (CSHQ) \citep{owens_childrens_2000} is an established, comprehensive method of assessing the degree of sleep difficulties in children. Responses across multiple domains are combined into a single total sleep disturbance score, with higher values indicating more severe problems and a conventional cut‑off used to flag likely clinical cases. This cumulative scoring framework is practical for screening but compresses heterogeneous sleep presentations into one dimension, potentially obscuring distinct profiles of difficulty that may warrant different intervention strategies. Furthermore, external factors such as age, gender and diagnosis of Autism Spectrum Disorder (ASD) may have a substantial impact on sleep difficulty and there is an unrealised opportunity to use this information to gain more insights on a child's sleep challenges and the factors impacting sleep. Modern latent class analysis (LCA)~\citep{goodman_exploratory_1974} approaches offer a principled framework for addressing these limitations by identifying distinct sleep profiles, while latent class regression (LCR) models~\citep{bandeen-roche_latent_1997,bolck_estimating_2004} can account for predictive factors. Recent methodological advances, including work by \citet{malsiner-walli_without_2025}, have expanded the scope and flexibility of LCA frameworks through mixture-of-mixtures approaches. Identifiability conditions for latent class models have been established by \citet{liu_restricted_2024} for restricted LCA models and by \citet{ouyang_identifiability_2022} for models with covariates.

In this paper we apply fully Bayesian LCA~\citep{garrett_latent_2000,white_bayesian_2016} and LCR models to a cohort of $N=148$ Irish children whose parent completed a CSHQ. As well as identifying distinctive sleep health profiles, our approach jointly performs item and covariate selection to determine which variables are relevant to the cluster analysis and which external factors are associated with specific sleep difficulties, respectively. 

Our model identifies one cluster of healthy sleepers and three distinct clusters of children presenting with problematic sleep patterns, while reducing the number of variables needed to cluster children from 33 to 21. We also find that ASD diagnosis is positively associated with some but not all problematic sleep clusters, in comparison to healthy sleepers. This analysis, leading to the differentiation of distinct sleep difficulties, contrasted with a more traditional identification of a paediatric sleep disorder is important. Specifically, a total sleep disturbance index on the CSHQ will lead a clinician to determine if a child presents with a sleep disorder but does not clearly indicate the specific subtype of associated sleep difficulties. In contrast, identifying a subtype of sleep disorder delineates a more specific, personalised and targeted pathway to interventions, rather than generic recommendations for children with sleep difficulties, particularly for children with an ASD diagnosis.

Our fully Bayesian model allows us to substantively address all of the key challenges in describing sleep health from CSHQ survey data. As well as determining the number of sleep difficulty profiles, our approach carries out covariate selection to determine predictors of those profiles. While principled model and item selection is well-developed for LCA, a novel contribution of our work is the development of a Bayesian LCR method that jointly performs predictor variable selection and item selection within the same algorithm, thus providing principled uncertainty quantification. We employ the Pólya-Gamma augmentation method \citep{polson_bayesian_2013} to enable efficient LCR model estimation and then show how this data augmentation technique can be extended to facilitate variable selection for both item variables and profile predictor covariates.  Covariate uncertainty is propagated by introducing latent covariate inclusion indicator variables \citep{holmes_bayesian_2006}, while a partially collapsed approach akin to that of \citet{white_bayesian_2016} is applied to allow for selection of item variables.

\subsection{Sleep Patterns Dataset}\label{subsec:sleep_data}

The Children’s Sleep Habits Questionnaire (CSHQ) is a widely utilised parent-report screening instrument for assessing sleep behaviours and identifying sleep difficulties in children. Developed by \citet{owens_childrens_2000}, it is commonly used for paediatric sleep evaluation in research and clinical settings. The CSHQ is a 35-item questionnaire in which parents retrospectively rate a recent typical week using a 3-point Likert scale: "usually" (5–7 times within the past week), "sometimes" (2–4 times), and "rarely" (never or 1 time). Selected items are reverse-scored so higher scores consistently indicate problematic sleep behaviours. Eight subscales correspond to common paediatric sleep issues — bedtime resistance (BR), sleep onset delay (SOD), sleep duration (SD), sleep anxiety (SA), night wakings (NW), parasomnias (P), sleep-disordered breathing (SDB), and daytime sleepiness (DS) — collectively assessing the frequency of prevalent difficulties. While CSHQ responses are ordinal, we treat them as nominal categorical variables for the purposes of LCA. Because BR4 (needs a parent in the room to sleep) and BR6 (afraid of sleeping alone) also appear under SA, these two responses are usually dropped for analysis, leaving 33 items. Summed responses yield a Total Sleep Disturbance index (range 33–99); a threshold of 41, established by \citet{owens_childrens_2000} via receiver operating characteristic (ROC) curve analysis, indicates a likely paediatric sleep disorder when exceeded. The CSHQ’s comprehensive assessment of sleep behaviours makes it valuable for identifying sleep problems in children, offering clinicians and researchers a standardised tool for screening and monitoring paediatric sleep difficulties.

The dataset analysed comprised CSHQ responses from 163 participants, a convenience sample of parents/caregivers and behaviour analysts within the Republic of Ireland. Eligibility required being a parent/caregiver living in the Republic of Ireland with a child aged 4–17 years, or a practising behaviour analyst in the Republic of Ireland. Parents and caregivers were recruited via social media parenting groups and by contacting primary and secondary schools; behaviour analysts were recruited via social media and registries of practising behaviour analysts in the Republic of Ireland. Of the 163 responses, 15 contained no survey item responses and were removed from further analysis. For a description of recorded demographic and diagnostic information, we refer to Table S6.

This analysis represents a secondary analysis utilising data from a previously conducted study forming part of an unpublished master’s thesis on barriers to accessing and successfully implementing behavioural sleep interventions in children with and without neurodevelopmental disabilities \citep{walsh_exploring_2022}. The previous analysis identified 113 participants as meeting criteria for a sleep disorder based on the conventional threshold $(>41)$, with a mean cumulative score of 47.28 for the full sample. An independent samples $t$-test indicated a significant group difference, with higher total CSHQ scores in the neurodiverse group (mean = 51.86, sd = 8.81) than in the neurotypical group (mean = 45.33, sd = 7.99), two-tailed $(t(146) = -4.40, p < 0.001)$. In addition to sleep behaviour data, demographic variables recorded included child age, family size, and gender.

While the CSHQ's cumulative scoring approach and established threshold of 41 provide a practical screening tool, this analytical strategy obscures important heterogeneity in paediatric sleep difficulties. Children with identical total scores may exhibit entirely different profiles of sleep disturbance across the eight subscales. For example, one child may struggle primarily with bedtime resistance and sleep onset delay, while another experiences sleep disordered breathing and daytime sleepiness, despite both exceeding the clinical cutoff. This presents a fundamental challenge for intervention: different patterns of sleep difficulty require different clinical approaches, yet the cumulative score cannot distinguish between them. Moreover, the binary classification imposed by the threshold reduces a multidimensional spectrum of sleep behaviours to a dichotomous outcome, masking gradations in severity and patterns of co-occurring difficulties. These limitations motivate the use of model-based approaches such as latent class analysis (LCA) and latent class regression (LCR), which can identify naturally occurring subgroups with distinct sleep profiles while simultaneously incorporating predictors of group membership and addressing which items are most informative for characterising heterogeneity.

\section{Methods}
\subsection{Latent Class Analysis and Latent Class Regression Models}\label{subsec:LCA_LCR}
Let $\boldY$ denote a matrix of dimension $N \times M$ where each row contains responses on $M$ categorical variables. The $i$-th row of $\boldY$ is given by $\boldy_i = (y_{i1}, y_{i2}, \dots, y_{iM})$. The number of possible responses for categorical variable $j$ is denoted by $K_j$.
Under the latent class analysis (LCA) model, the response vector $\boldy_i$ is assumed to arise from one of $G$ latent classes, with class probabilities $\bspi$. 
Therefore the observed-data likelihood for observation $i$ is
\begin{equation}\label{mixture_likelihood}
    p(\boldy_i \mid \bspi, \bsthet) = \sum_{g=1}^G \pi_g p(\boldy_i \mid \bsthet_g),
\end{equation}
where $\bsthet_{gj}$ are the parameters that define the item response probabilities for the variables $j = 1,\dots,M$ conditional on membership in group $g$. Specifically, given that an observation belongs to group $g$, we define $\theta_{gjk}$ as the probability of response $k$ for item $j$. In the standard LCA model, each of the item variables are assumed to be locally independent given group membership. Hence we can write the likelihood for an observation $\boldy_i$ conditional on group membership as
\[
p(\boldy_i \mid \bsthet_g) = \prod_{j=1}^{M}\prod_{k=1}^{K_j}\theta_{gjk}^{\,\delta_k(y_{ij})},%I(y_{ij} = k)}.
\]
where $\delta_k(y) = 1$ if $y=k$ and 0 otherwise. The LCA model can be extended to incorporate concomitant variables --- covariates that influence the prior probability of class membership. This extended model is commonly referred to as the latent class regression (LCR) model. Let $\boldX$ denote a regression matrix of dimension $N \times (P+1)$, where $P$ is the number of covariates, and each row is given by $\boldx_i = (1, x_{i1},\dots,x_{iP})^T$, with $x_{il}$ denoting the value of the $l$-th covariate for observation $i$, where $l = 1,\dots,P$. The prior probability that observation $i$ belongs to group $g$ is then modelled using a softmax (multinomial logit) function. This involves setting 
\[
\pi_{ig} = \frac{\exp\left( \boldx_i^T \bsbet_g \right)}{\sum_{h=1}^G \exp\left( \boldx_i \bsbet_h \right)},
\]
where $\bsbet_g = (\beta_{g0}, \beta_{g1},\dots,\beta_{gP})^T$ is the vector of regression coefficients for membership in group $g$. The observed-data likelihood for $\boldy_i$ given $\bsbet, \bsthet$ is obtained by replacing the mixing parameters $\pi_g$ in (\ref{mixture_likelihood}) with the observation-specific probabilities $\pi_{ig}$. One of the groups is selected as the baseline against which the log-odds of class membership are compared. In our case, we choose this group to be $G$, i.e. we set $\bsbet_G = \mathbf{0}$. Since it is difficult to carry out inference using the observed-data likelihood (\ref{mixture_likelihood}), it is useful to introduce missing latent class labels and work with a complete-data likelihood instead. For an observation $i$, the class label vector $\boldz_i \in \{0,1\}^{G}$ is given by
\[
z_{ig} = \begin{cases}
    1,\; \text{ if observation $i$ belongs to group $g$,}\\
    0,\; \text{ otherwise.}
\end{cases}
\]
The complete data likelihood for $\boldy_i$ and $\boldz_i$ under the LCR model is then given by
\begin{equation}\label{LCR_complete_like}
p(\boldy_i, \boldz_i \mid \bsbet, \boldx_i, \bsthet) = \prod_{g=1}^G \left[ \pi_{ig} \prod_{j=1}^M \prod_{k=1}^{K_j} \theta_{gjk}^{\,\delta_k(y_{ij})} \right]^{z_{ig}}.%I(y_{ij} = k)
\end{equation}
The standard LCA model likelihood is recovered by marginalising over the predictor variables, as shown by \citet{bandeen-roche_latent_1997}. One can view the model as a finite mixture of experts (MoE) (\citet{jacobs_adaptive_1991}; \citet{gormley_mixture_2019}), where each latent class defines an expert characterised by its item‑response distribution, and a regression on predictors governs the class‑membership probabilities (mixing weights).
\subsection{Estimation Approaches}\label{subsec:lca_gibbs}
When $G$ is fixed, several approaches are available for estimating the LCA model parameters. Maximum likelihood typically uses the Expectation-Maximisation (EM) algorithm \citep{dempster_maximum_1977}. In a Bayesian framework, choosing standard conjugate priors allows for the full conditional distributions of the parameters $\bsthet, \bspi$ and $\boldZ$ to be derived in closed form, making Gibbs sampling \citep{gelfand_sampling-based_1990} a natural choice for posterior inference. We give a brief outline of the Gibbs sampling method developed by \citet{garrett_latent_2000}. Let $\Dir(\bsphi)$ denote a Dirichlet density on the $d-1$ simplex with parameter $\bsphi = (\phi_1,\dots,\phi_d)$. We can impose priors $\bsthet_{gj} \sim \Dir(\bsal_j)$, $\bspi \sim \Dir(\bslam)$, where $\bsal_j = (\alpha_1, \dots, \alpha_{K_j})$ and $\bslam = (\lambda_1,\dots,\lambda_G)$. Denoting $\boldS_{gj} = (s_{gj1}, \ldots, s_{gjK_j})$ and $\boldS = (s_1, \ldots, s_G)$, the full conditional distributions of the relevant parameters to be sampled iteratively in a Gibbs sampler are
\begin{eqnarray*}
\bsthet_{gj} &\mid& \boldY, \boldZ \sim \Dir(\bsal_j + \boldS_{gj})\\
\bspi &\mid & \boldZ \sim \Dir(\bslam + \boldS)\\
\boldZ_i &\mid& \bsthet, \bspi, \boldY_i \sim \Mult(1,\boldw_i)
\end{eqnarray*}
where 
\begin{equation}\label{LCA_sampling_quantities}
w_{ig} \propto \pi_g \prod_{j=1}^M \prod_{k=1}^{K_j} \theta_{gjk}^{\,\delta_k(y_{ij})},\qquad s_{gjk} = \sum_{i=1}^N z_{ig}\,\delta_k(y_{ij}), \qquad s_g = \sum_{i=1}^N z_{ig}. %I(y_{ij} = k)
\end{equation}
In carrying out maximum likelihood estimation for an LCR model, an EM algorithm can be applied using either the one-step or three-step methods. The three-step (\citet{vermunt_latent_2010}; \citet{bolck_estimating_2004}) involves fitting a standard LCA model, followed by assigning each observation to one of the fitted groups based on a chosen assignment rule. A multinomial logistic regression is then fit with the group labels as response variables. 

The one-step EM method \citep{bandeen-roche_latent_1997} directly includes  covariates and estimates the regression coefficients simultaneously.  
\subsection{Bayesian LCR Model with Pólya-Gamma Augmentation}
For a Bayesian formulation of an LCR model, posterior inference involves sampling from the joint posterior of the latent class labels $\{\boldz_i\}_{i=1,\dots,N}$, the class-specific response probabilities $\{\bsthet_{gj}\}_{j=1,\dots,M}^{g=1\dots,G}$, and regression coefficients $\{\bsbet_g\}_{g = 1,\dots,G-1}$. With conjugate priors selected for $\bsthet_{gj}$, sampling $\boldz_{i}$ and $\bsthet_{gj}$ follows the standard LCA approach. As outlined in Section \ref{subsec:lca_gibbs}, setting $\bsthet_{gj} \sim \Dir(\bsal_j)$ yields the full conditional $\bsthet_{gj} \mid \boldY, \boldZ \sim \Dir(\bsal_j + \boldS_{gj})$ as before. Similarly, the class membership labels follow $\boldz_i \mid \bsthet, \boldx_i, \{\bsbet_{g}\}_{g=1,\dots,G}, \boldy_i \sim \Mult(1,\boldw_i)$, where 
\[
w_{ig} \propto \frac{\exp\left( \boldx_i^T \bsbet_g \right)}{\sum_{h=1}^G \exp\left( \boldx_i \bsbet_h \right)} \prod_{j=1}^M \prod_{k=1}^{K_j} \theta_{gjk}^{\,\delta_k(y_{ij})} .%{I(y_{ij} = k)}.
\]
We encounter difficulties however when we attempt to draw samples from the full conditional of the $\bsbet_g$ parameters. Let $\bsbet$ denote the $(P+1)\times G$ dimensional matrix whose columns are the regression coefficient vectors $\bsbet_g$, $g = 1,\dots, G$.
Considering some prior distribution $p(\bsbet)$, with $\bsbet_G = \mathbf{0}$ fixed for identifiability, the form of the full conditional distribution for $\bsbet$ is given by
\begin{equation}\label{beta_fc1}
p(\bsbet \mid \boldX, \boldZ) \propto p(\bsbet)\, \prod_{i=1}^N \prod_{g=1}^G \left[ \frac{\exp(\boldx_i^T\bsbet_g)}{\sum_{h=1}^G\exp(\boldx_i^T\bsbet_h)} \right]^{z_{ig}}.
\end{equation}
There is no conjugate choice for $p(\bsbet)$ in this situation, so direct Gibbs sampling of the $\bsbet_g$ using full conditionals is not possible. One approach to overcome this barrier is to use a Metropolis-within-Gibbs step to sample the coefficient parameters, which is employed by \citet{chung_latent_2006}. However, Metropolis-within-Gibbs approaches become computationally prohibitive for large numbers of groups or predictor variables due to slow mixing and the need for extensive tuning of proposal distributions.

Pólya-Gamma (PG) augmentation \citet{polson_bayesian_2013} has emerged as a flexible alternative for Bayesian inference in binary and multinomial logistic regression models. This data augmentation technique introduces auxiliary variables which are PG distributed, and which render the logistic likelihood conditionally Gaussian, facilitating Gibbs sampling of regression coefficients. Recent advances in PG augmentation include the Ultimate Pólya-Gamma (UPG) by \citet{zens_ultimate_2024}, which is applied in the context of binary and multinomial logistic regression and further improves computational efficiency of the PG sampler. 

They also outline an MoE extension as an application of the improved sampler. While highly effective for regression settings, the emphasis of their work is on computational properties of the sampler and MCMC performance. Their UPG sampler has not been applied in the context of the LCR model. Furthermore, it does not address the task of estimating uncertainty in the structure of the model. The UPG sampling approach has also been extended to other mixture modelling contexts, including an inference approach for censored MoE models \citep{mirfarah_robust_2024}. Beyond regression settings, PG augmentation has also been used to enable efficient Gibbs sampling in latent factor models for multivariate binomial data \citep{holmes_polya-gamma_2022}.
\subsubsection{Pólya-Gamma Augmentation}\label{subsec:pg_augmentation} 
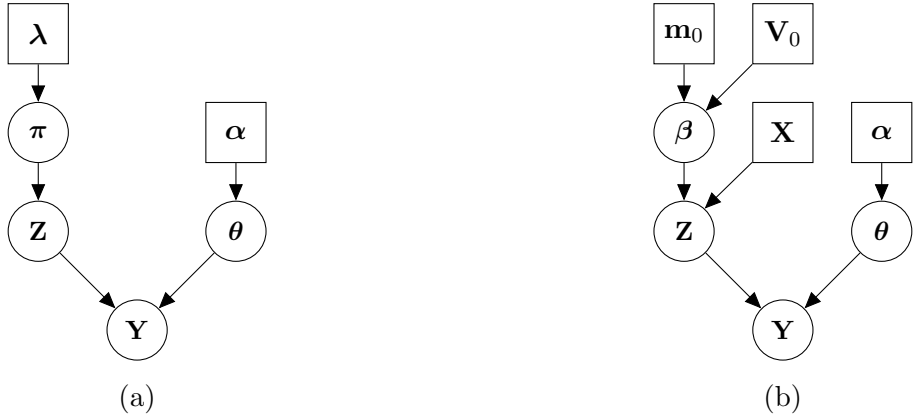
\begin{figure}[htbp]
    \centering
    \begin{minipage}{0.48\textwidth}
        \centering
        \begin{tikzpicture}[scale=0.85, every node/.style={scale=1}]
        \matrix[column sep=0.5cm, row sep=0.5cm] {
        \node[draw, rectangle, minimum width=0.8cm, minimum height=0.8cm] (d) {$\bslam$}; & &\\
        \node[latent, minimum width=0.8cm, minimum height=0.8cm] (p) {$\bspi$}; & &
        \node[draw, rectangle, minimum width=0.8cm, minimum height=0.8cm] (a) {$\boldsymbol{\alpha}$}; \\
        \node[latent, minimum width=0.8cm, minimum height=0.8cm] (z) {$\boldZ$}; &
        &
        \node[latent, minimum width=0.8cm, minimum height=0.8cm] (t) {$\boldsymbol{\theta}$}; \\
        &
        \node[latent, minimum width=0.8cm, minimum height=0.8cm] (y) {$\boldY$}; &
        \\
    };

    \draw[->] (d) -- (p);
    \draw[->] (p) -- (z);       
    \draw[->] (a) -- (t);      
    \draw[->] (z) -- (y);       
    \draw[->] (t) -- (y);
        \end{tikzpicture}
        
        (a)
    \end{minipage}
    \hfill
    \begin{minipage}{0.48\textwidth}
        \centering
        \begin{tikzpicture}[scale=0.85, every node/.style={scale=1}]
        \matrix[column sep=0.5cm, row sep=0.5cm] {
        \node[draw, rectangle, minimum width=0.8cm, minimum height=0.8cm] (m0) {$\boldm_0$};& \node[draw, rectangle, minimum width=0.8cm, minimum height=0.8cm] (V0) {$\boldV_0$};&\\
        \node[latent, minimum width=0.8cm, minimum height=0.8cm] (b) {$\boldsymbol{\beta}$}; &
        \node[draw, rectangle, minimum width=0.8cm, minimum height=0.8cm] (x) {$\boldX$}; &
        \node[draw, rectangle, minimum width=0.8cm, minimum height=0.8cm] (a) {$\boldsymbol{\alpha}$}; \\
        \node[latent, minimum width=0.8cm, minimum height=0.8cm] (z) {$\boldZ$}; & &
        \node[latent, minimum width=0.8cm, minimum height=0.8cm] (t) {$\boldsymbol{\theta}$}; \\
        &
        \node[latent, minimum width=0.8cm, minimum height=0.8cm] (y) {$\boldY$}; &
        \\
    };

    \draw[->] (m0) -- (b);
    \draw[->] (V0) -- (b);
    \draw[->] (x) -- (z);       
    \draw[->] (b) -- (z);       
    \draw[->] (a) -- (t);       
    \draw[->] (z) -- (y);       
    \draw[->] (t) -- (y); 
        \end{tikzpicture}
        
        (b)
    \end{minipage}
    \caption{Dependency graphs for (a) the LCA model and (b) the LCR model, with Pólya-Gamma data augmentation. }
     \label{fig:graph}
\end{figure}

The Pólya-Gamma data augmentation approach of \citet{polson_bayesian_2013} facilitates Bayesian inference in logistic-type models by introducing latent variables that give the logit likelihood a Gaussian form.
The primary result presented by \citet{polson_bayesian_2013} is the following. Let $\omega \sim \pg(b,0)$ for $b>0$. Then the following identity holds for any $a \in \mathbf{R}$:
\begin{equation}\label{pgresult}
\frac{(e^{\psi})^a}{(1+e^{\psi})^b}  
    = 2^{-b} e^{\kappa \psi} \Ex_{\omega}\left[ \exp\left(-\frac{\omega \psi^2}{2}\right) \right],
\end{equation}
where $\kappa = a-\frac{b}{2}$. Furthermore, we also have that $\omega \mid \psi \sim \pg(b,\psi)$. For a full proof of this result, we refer to \cite{polson_bayesian_2013}. Assuming independent priors $\bsbet_g \sim \cN(\boldm_{0g}, \boldV_{0g})$, $g = 1,\dots,G-1$ and applying this result to the LCR model, we obtain a two-step sampling method for $\bsbet_g$ and $\omega_{ig}$, $i = 1,\dots, N$, $g = 1, \dots, G-1$, given by
\begin{equation}\label{LCR_beta_sampling}
 \begin{aligned}
    \bsbet_g \mid \bsbet_{-g}, \omega_{\cdot g}, \boldX,\boldZ  &\sim \cN(\mathbf{m}_g, \mathbf{V}_g),\\
    \omega_{ig} \mid \bsbet, \boldX_i &\sim \pg(1,\eta_{ig}).
\end{aligned}
\end{equation}
where
\[
\eta_{ig} = \boldx_i^T\bsbet_g - c_{ig}, \qquad c_{ig} = \log\left( \sum_{h \neq g} \exp(\boldx_i^T \bsbet_h) \right), \qquad \mathbf{V}_g^{-1} = \boldX^T \Omega_g \boldX + \boldV_{0g}^{-1},
\]
\[
\boldm_g = \boldV_g \left( \boldX^T(\bskap_g+ \Omega_g\boldC_{\cdot g}) + \boldV_{0g}^{-1}\boldm_{0g} \right), \qquad \Omega_g = \diag(\{\omega_{ig}\}_{i=1}^n), \qquad \kappa_{ig} = z_{ig} - \frac{1}{2},
\]
$\boldC_{\cdot g}$ is the $g$-th column of $\boldC = (c_{ig})$, $\boldm_{0g}$ and $\boldV_{0g}$ are the prior mean and covariance of $\bsbet_g$, respectively. For the full derivation, see Section S3. Efficient sampling from a Pólya-Gamma distribution is facilitated by the \verb|rpg| function, included in the R package \verb|BayesLogit| \citep{polson_bayesian_2013}. Graphical diagrams for the LCA and LCR models are shown in Figure~\ref{fig:graph}.
\subsection{Variable Selection Extensions}
The number of groups $G$ that are present within the data are usually unknown and should be inferred. Another important consideration to make when performing LCA is that of item variable selection, since including variables that don't distinguish groups can impact the performance and interpretability of clustering. The literature on variable and model selection for LCA includes methods both from the frequentist and the Bayesian paradigm. For a detailed overview of model and variable selection methods for LCA, we refer to \citet{fop_variable_2018}. 
\citet{pan_bayesian_2014} propose a Bayesian LCR framework that jointly estimates the number of groups $G$ and model parameters via RJMCMC. However, it lacks variable selection, and relies on rejection sampling of regression parameters, which can be inefficient in high dimensions.
\citet{white_bayesian_2016} present a fully Bayesian collapsed Gibbs approach for LCA that treats variable and model selection as a probabilistic search over clustering solutions and variable subsets. However, this does not allow for predictor variables and does not account for redundancy between items.
The stepwise selection strategy of \citet{fop_variable_2017} is designed to exclude both non‑informative and redundant variables, and has recently been extended by \citet{xu_variable_2024} to handle missing data in LCA. However this method doesn't allow for predictor variable selection within LCR. 
Despite significant advances in variable selection for LCA, there remains a gap in the context of the LCR model. Currently, there are no widely adopted, dedicated methods that jointly select both item and predictor variables. We extend the Gibbs sampling method to include variable selection for both item variables and predictor variables. 
\subsubsection{Item Variable Selection}\label{subsubsec:item_var_sel}
In order to apply item variable selection, a partial collapsing approach similar to that of \citet{white_bayesian_2016} can be employed. Let $\bsnu_{\cl}$ denote the set of item variables that are useful for clustering, and let $\bsnu_{\nsub}$ be the set of variables that are not useful. We write the entire set of possible item variables as $\bsnu = (\bsnu_{\cl}, \bsnu_{\nsub})$. The responses to variables contained in $\bsnu_{\cl}$ are assumed to be produced by an LCR model with $G$ classes. Hence the complete data likelihood for these variables can be expressed as before,
\begin{equation}\label{clust_var_likelihood}
    p_{\cl}(\boldy_i, \boldz_{i} \mid \boldx_i, \bsbet, \bsthet, \bsnu) = \prod_{g=1}^G \left[ 
    \pi_{ig} \prod_{j \in \bsnu_{\cl}} \prod_{k=1}^{K_j} \theta_{gjk}^{\,\delta_k(y_{ij})} \right]^{z_{ig}}. 
\end{equation}
The responses to variables in $\bsnu_{\nsub}$ are assumed to be produced by an LCA model with only one class. By denoting the probability of a non-clustering variable $j$ taking response $k$, the likelihood can be written as
\begin{equation}\label{nonclust_var_likelihood}
p_{\nsub}(\boldy_i \mid \bsrho, \bsnu) = \prod_{j \in \bsnu_{\nsub}} \prod_{k=1}^{K_j} \rho_{jk}^{\delta_k(y_{ij})}. %I(y_{ij} = k)
\end{equation}
Then the complete data likelihood $p(\boldy_i, \boldz_i \mid \boldx_i, \bsbet, \bsthet, \bsrho, \bsnu)$ for all variables is given by the product of (\ref{clust_var_likelihood}) and (\ref{nonclust_var_likelihood}). 
The joint posterior of the parameters of interest is given by
\begin{equation}\label{varsel_posterior}
    p(\boldZ, \bsthet, \bsrho, \bsnu, \bsbet \mid \boldX, \boldY) \propto p(\boldY, \boldZ \mid \boldX, \bsbet, \bsthet, \bsrho, \bsnu)\,p(\bsthet)\,p(\bsrho)\,p(\bsbet)\,p(\bsnu),
\end{equation}
where the prior distributions for $\bsthet$, $\bsrho$ and $\bsnu$ are the same as those detailed in \cite{white_bayesian_2016}, and the prior distribution for $\bsbet_g$ is the same as in Section \ref{subsec:pg_augmentation}.
We cannot estimate the uncertainty in the number of groups in this situation without the need for dimension jumping adjustments, due to the presence of the regression parameters. However, it is still possible to marginalise with respect to the item probability parameters, which allows us to quantify the clustering usefulness of each of the item variables. We analytically integrate expression (\ref{varsel_posterior}) with respect to the $\bsthet$ and $\bsrho$ parameters,
which gives 
\begin{multline}\label{varsel_likelihood}
p(\boldZ, \bsnu, \bsbet \mid \boldX, \boldY) \propto p(\bsbet)\,p(\bsnu)\,\prod_{i=1}^N \prod_{g=1}^G 
\pi_{ig}^{z_{ig}} \left[ \prod_{j\in \bsnu_{\nsub}} \frac{\Gamma(K_j \alpha)}{\Gamma(\alpha)^{K_j}} \frac{\prod_{k=1}^{K_j} \Gamma(s_{jk} + \alpha)}{\Gamma(N + K_j\alpha)}\right]\\ \times \left[ \prod_{g=1}^G \prod_{j \in \bsnu_{\cl}} \frac{\Gamma(K_j \alpha)}{\Gamma(\alpha)^{K_j}}\frac{\prod_{k=1}^{K_j}\Gamma(s_{gjk} + \alpha)}{\Gamma(s_g + \alpha)} \right].
\end{multline}
Here, $s_g$ and $s_{gjk}$ are as in (\ref{LCA_sampling_quantities}), and $s_{jk} = \sum_{i=1}^n \delta_k(y_{ij})$. The sampling method for item variable selection closely follows the approach of \cite{white_bayesian_2016}. For the logit coefficient parameters $\bsbet_g$, the method is the same as previously. Since the full conditional for $\bsbet$ will take the same form as (\ref{beta_fc1}), we get the same form for the Gaussian that $\bsbet_g$ will be drawn from, namely (\ref{LCR_beta_sampling}). For the inclusion/exclusion of the clustering variables, a Metropolis step is employed. The method proceeds as follows: An index $j$ is selected at random among the indices of each of the variables $1,\dots,M$. If the variable $j \in \bsnu_{\cl}$ currently, then it is proposed that it be moved to $\bsnu_{\nsub}$. Similarly, if $j \in \bsnu_{\nsub}$, then a proposal is made to move it to $\bsnu_{\cl}$. In the case that the chosen $j \in \bsnu_{\nsub}$, then the acceptance probability of $j$ being moved to $\bsnu_{\cl}$ is given by $\min(1,A)$, where
\[
    A = \frac{p(\boldZ, \bsnu^*, \bsbet \mid \boldX, \boldY)}{p(\boldZ, \bsnu, \bsbet \mid \boldX, \boldY)}.
\]
In the case where the proposed variable $j \in \bsnu_{\cl}$, the probability of acceptance for the move to $\bsnu_{\nsub}$ is given by $\min(1,A^{-1})$. To sample the group labels $\boldZ_i$, we can remove observation $i$ from its current group assignment, followed by assigning it to each of the groups $h = 1,\dots,G$ and evaluating the likelihood expression (\ref{varsel_likelihood}) for each of the assignments. 
The new class assignment $\boldz_i^*$ is drawn from a $\Mult(1,\boldw_i)$ distribution, where $\boldw_i$ is the normalised value of the vector of likelihood evaluations. This is repeated for each observation $i$ on every iteration.
\subsubsection{Predictor Variable Selection}\label{subsec:pred_sel}
For selection of predictor variables, we utilise a method similar to that of \citet{holmes_bayesian_2006}, which has been applied by \citet{dempsey_bayesian_2025} in an application to distributed lag models. We introduce the latent variables $\gamma_l \in \{0,1\}$ for each predictor variable $\boldX_{\cdot l}$, $l = 1,\dots,P$. These variables are indicators which determine whether the corresponding covariate is included in the model. The vector $\bsgam$ can be updated by using a Metropolis step with a joint proposal with $\bsbet$. We have the following joint proposal decomposition,
\[
q(\bsbet^*, \bsgam^*) = \pi(\bsbet^* \mid \bsgam^*, \bsom, \boldX, \boldZ)q(\bsgam^* \mid \bsgam).
\]
The first factor in this product is the previously derived full conditional distribution of $\bsbet$ from (\ref{LCR_beta_sampling}), but with the parameters $\boldm_g$, $\boldV_g$ computed using only the columns $\boldX_{\cdot l}$ corresponding to $\gamma_{l} = 1$. The second factor is some proposal method for $\bsgam$. We can factor the joint posterior for $\bsbet$ and $\bsgam$ as
\[
\pi(\bsgam, \bsbet \mid  \bsom, \boldX, \boldZ) = \pi(\bsgam \mid \bsom, \boldX, \boldZ)\pi(\bsbet \mid \bsgam, \bsom, \boldX, \boldZ).
\]
The acceptance criterion for the inclusion or exclusion of a predictor on a given iteration is then given by 
\[
   \alpha = \min \left( 1, \frac{|\boldV_{\bsgam^*}|^{\frac{1}{2}}|\boldv_{\bsgam}|^{\frac{1}{2}}\exp\left( \frac{1}{2} \left( \boldm_{\bsgam^*}^T\boldV_{\bsgam^*}\boldm_{\bsgam^*} + \boldm_{0\bsgam^*}^T\boldV_{0\bsgam^*}^{-1}\boldm_{0\bsgam^*} \right) \right)q(\bsgam \mid \bsgam^*)\pi(\bsgam^*)}
   {|\boldV_{\bsgam}|^{\frac{1}{2}}|\boldv_{\bsgam^*}|^{\frac{1}{2}}\exp\left( \frac{1}{2} \left( \boldm_{\bsgam}^T\boldV_{\bsgam}\boldm_{\bsgam} + \boldm_{0\bsgam}^T\boldV_{0\bsgam}^{-1}\boldm_{0\bsgam} \right) \right)q(\bsgam^* \mid \bsgam)\pi(\bsgam)} \right) .
\]
The mean vectors $\boldm_{\bsgam}$ are calculated by concatenating the means $\boldm_{g, \bsgam}$, and in a similar way, the covariance matrix $\boldV_{\bsgam}$ is the block diagonal matrix containing the matrices $\boldV_{g, \bsgam}$ as the diagonal elements. The subscript $\bsgam$ in the mean $\boldm_{\bsgam}$ and covariance $\boldV_{\bsgam}$ indicates that the parameters are computed with only the columns of $\boldX$ supported by the configuration $\bsgam$. Since $\bsbet$ doesn't appear in the acceptance criterion, there is no need for dimension jumping adjustments in the expression. The $\bsbet$ parameters can then be updated through the previously described Gibbs step conditional on the current configuration of $\bsgam$. Regarding the proposal density $q$, in this case a predictor was proposed for inclusion/exclusion by uniformly selecting one of the indices. Since this is a symmetric proposal, there is no contribution from $q$ in the acceptance criterion. The prior $\pi(\bsgam)$ was set to be Bernoulli with success parameter $0.5$. 
\subsection{Post-hoc processing of samples}
In order to perform inference and interpret the MCMC output, a number of post-processing steps must be taken.

\subsubsection{Sample relabelling}
The well known-label switching problem for mixture models \citep{celeux_computational_2000} must be handled appropriately for this model.
Methods for post-hoc relabelling of MCMC output have been well established. We used the assignment algorithm of \citet{stephens_dealing_2000} to post-process samples, using the R package \verb|label.switching| \citep{papastamoulis_labelswitching_2016}. 
A second related issue arises during the relabelling process for the LCR model. The coefficient parameters $\bsbet_g$ are defined relative to a baseline group, which we have designated as group $G$, such that $\bsbet_G = \mathbf{0}$. However, if the labels switch during the sampling process, so that the group originally designated as $G$ is reassigned to a different group index $h$, then all $\bsbet_g$ parameters become redefined relative to a new baseline, specifically the group now indexed as $G$. As a result, following such a label switch, the samples of $\bsbet_g$ no longer correspond to a consistent baseline, and are instead drawn from distributions that are defined relative to whichever group is currently labelled as $G$ in each iteration. Therefore, following the relabelling procedure, the relevelled coefficient sample for group $g$ at iteration $t$, denoted by ${\bsbet'}_g^{(t)}$, is obtained by shifting the original sample relative to the baseline group $G$, i.e., ${\bsbet'}_g^{(t)} = \bsbet_g^{(t)} - \bsbet_G^{(t)}$.

\subsubsection{Post-hoc point estimation of clustering partition and parameters}\label{subsubsec:clust_point_est}
Assigning observations to groups in LCA enables interpretation of latent structure and examination of group characteristics. While LCA Gibbs samplers produce soft classifications via posterior assignment probabilities, computing a clustering point estimate is useful for analyzing model structure and summarizing group compositions.
The point estimation method we use in this case is the minimum Variation of Information (VI) method of \citet{wade_bayesian_2018}. This aims to find a cluster partition that minimises the posterior expected VI. We apply the greedy algorithm mentioned in \citet{wade_bayesian_2018} to obtain this optimal partition $\widehat{\boldc}$, which is implemented using the \verb|minVI| function included in the R package \verb|mcclust.ext| \citep{wade_mcclustext_2015}. In order to obtain uncertainty quantification, \citet{wade_bayesian_2018} also extend this point estimation method to credible balls for clustering solutions.  
The calculation of the upper and lower vertical bounds and the horizontal bounds of the $(1-\alpha)\%$ credible ball around $\boldc^*$ can be carried out using the \verb|credibleball| function in the R package \verb|mcclust.ext|. The item probability parameters $\bsthet$ are marginalised from the model in its collapsed form, so we cannot directly draw samples from their full conditionals. However, in the same way as was used in the collapsed sampler for the LCA model by \citet{white_bayesian_2016}, we can use standard results to compute the mean and variance of these parameters.
\subsection{Overall Inference Workflow and Prior Specification}\label{subsec:inferenceworkflow}
Our inference procedure uses a two-stage design. Since the regression coefficients $\bsbet_g$ preclude full model collapse in the LCR framework and thus prevent direct Bayesian estimation of $G$ during variable selection, we first estimate $G$ externally before conducting LCR fitting and variable selection. This approach is justified by \citet{bandeen-roche_latent_1997}, who demonstrate that  marginalising over covariates in the LCR likelihood yields the standard LCA likelihood, enabling recovery of clustering-relevant group structure independently of covariates. The workflow proceeds as follows: 
\begin{enumerate}
    \item The collapsed sampler of \citet{white_bayesian_2016} is applied to the LCA model, yielding posterior probabilities for plausible $G$ values and MCMC samples for cluster partitions. The number of groups is selected based on this posterior distribution while examining cluster point estimates and 95\% credible balls from the minimum posterior expected VI method to ensure meaningful clusters and avoid overfitting.
    \item With $G$ fixed at the selected value, the LCR model is fitted\footnote{While it did not occur in our analysis, if substantial posterior uncertainty in $G$ were to occur after Step 1, we recommend repeating Step 2 across all values with substantive support (posterior probability $> 0.10$) and presenting final conclusions along with a sensitivity analysis.}. We compare four model configurations in Step 2: no variable selection, item selection only, predictor variable selection only, and joint item and predictor variable selection. 
\end{enumerate}
Unless otherwise stated, the following configuration is used throughout Sections \ref{sec:sim_study}-\ref{sec:results}. MCMC chains are run for 50,000 iterations following 1000 burn-in iterations, retaining every 10th sample by thinning. Hyperparameters are specified to be weakly informative: $\bsal_j = \mathbf{1}_{K_j}$ for item probability parameters $\bsthet_{gj}$, $g = 1,\dots,G$; $\boldm_{0g} = \mathbf{0}$, $\boldV_{0g} = 10^2\boldI_{p+1}$ for regression coefficients $\bsbet_g$; and $\Bern(0.5)$ priors are chosen for predictor inclusion indicators $\bsgam_l$. For the collapsed model, independent prior probabilities of 0.5 are assumed for clustering variable utility, i.e. we use $p(\bsnu) = \prod_{j=1}^M p(\nu_j)$, a product of independent priors, where $p(\nu_j)$ is $\mathrm{Bernoulli}(0.5)$.
\section{Simulation Study}\label{sec:sim_study}
We run two simulation studies to assess the performance of the sampling method. Model performance was assessed in terms of parameter estimation, and model, item and predictor variable selection accuracy. Cluster accuracy was assessed using adjusted Rand index (ARI) \citep{hubert_comparing_1985}. In both cases, the datasets consisted of $N = 500$ observations. Predictor variable selection accuracy was also evaluated for in a range of scenario analyses for $N = 150$ to $N = 500$.
\subsection{Dataset 1}\label{subsec:sim_study1}
The first scenario simulates a straightforward two class structure. We consider $M = 8$ response items, each with three levels. The variables 1, 2, 3, 4 are discriminative, while 5, 6, 7, 8 do not vary across groups.  Table \ref{table:sim1_item_prob_estimates_full_LCR} presents the item response probabilities for each group.
For this simulation, we consider $P = 6$ possible independent predictor variables, 2 of which are noise. The rest of the variables have varying effect strength. The coefficient parameter vector for membership of Group 1 is given by $\bsbet = (-0.5, 1.2, 1.0, 0.8, 0.4, 0.0, 0.0)^T$. All covariates are simulated from independent $\cN(0,1)$ distributions.

\subsubsection{Results}
\begin{figure}
    \centering
    \includegraphics[width=\linewidth]{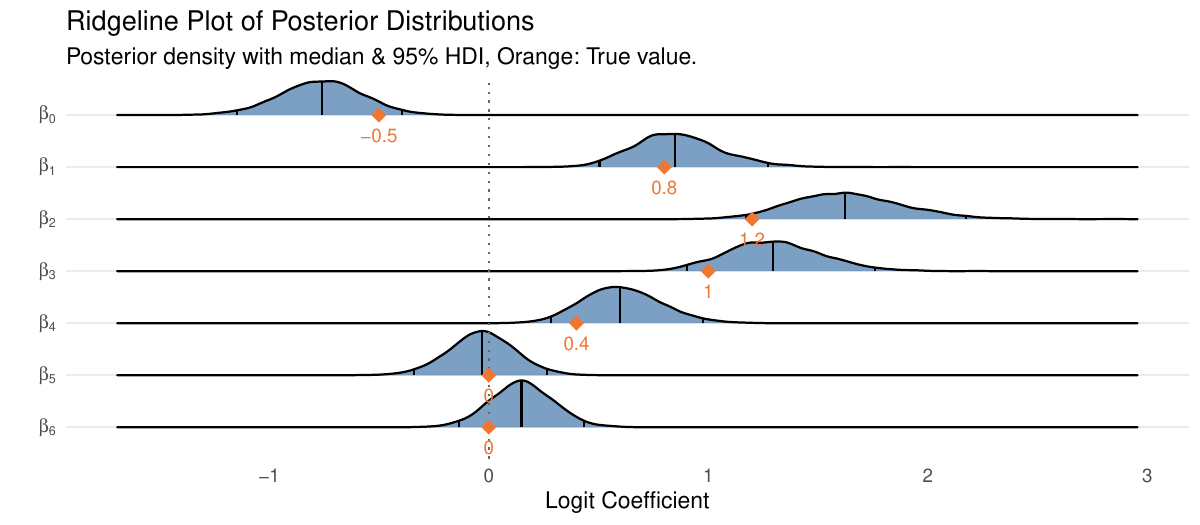}
    \caption{Ridgeline plots of posterior distribution of regression coefficients for Simulation 1.}
    \label{fig:sim1_ridgeline}
\end{figure}

Initially, the sampler was run for the full LCR model, without any variable selection. The model parameter estimates were accurate. Figure \ref{fig:sim1_ridgeline} contains a ridgeline plot which summarises the estimated posterior densities of the coefficient parameters, along with the 95\% highest density intervals estimated from the MCMC samples and the true simulated parameter values. In all cases, the true simulated parameter value was contained within the 95\% HDI. For item probability estimates, we refer to Table S1.
\begin{table}
    \centering
    \begin{tabular}{cccc}
    \toprule
    \multicolumn{4}{c}{\textbf{Discriminative Variables}}
    \\
    \midrule
    \textbf{Var.} & \textbf{Cat.} & \textbf{True 1} & \textbf{True 2}
    \\
    \midrule
    1 & 1 & 0.15  & 0.70     \\
  & 2 & 0.25  & 0.20    \\
  & 3 & 0.60  & 0.10    \\
2 & 1 & 0.20  & 0.55    \\
  & 2 & 0.35  & 0.30    \\
  & 3 & 0.45  & 0.15    \\
3 & 1 & 0.10  & 0.80    \\
  & 2 & 0.15  & 0.15     \\
  & 3 & 0.75  & 0.05     \\
4 & 1 & 0.25  & 0.45     \\
  & 2 & 0.40  & 0.35     \\
  & 3 & 0.35  & 0.20     \\
  \bottomrule
  \end{tabular}
  \hspace{2em}
  \begin{tabular}{ccc}
  \toprule
  \multicolumn{3}{c}{\textbf{Non-discriminative Variables}}
  \\
  \midrule
    \textbf{Var.} & \textbf{Cat.} & \textbf{True}
    \\
    \midrule
  5 & 1 & 0.40     \\
  & 2 & 0.50     \\
  & 3 & 0.10    \\
6 & 1 & 0.70     \\
  & 2 & 0.10     \\
  & 3 & 0.20    \\
7 & 1 & 0.10    \\
  & 2 & 0.50    \\
  & 3 & 0.40      \\  
8 & 1 & 0.33   \\
  & 2 & 0.33     \\
  & 3 & 0.33    \\ 
  \bottomrule
  \end{tabular}
    \caption{True values of the item probability parameters for Simulation 1.}
    \label{table:sim1_item_prob_estimates_full_LCR}
\end{table}
In the item selection run, items 1, 2, 3, 4 were correctly identified as being discriminative of classes, with sample estimated posterior probabilities $\widehat{\prob}(\nu_j = 1 \mid \text{everything else}) = 1.00$ for each item. In contrast, items 5, 6, 7, 8 had posterior inclusion probabilities less than 0.08, indicating strong evidence for their exclusion. We refer to Table S3 for the complete posterior inclusion probabilities of each item variable. The item probability parameters were computed post-hoc, and are provided in Table S2. For the item selection model, the estimates for the $\bsbet$ coefficient parameters did not differ substantially from the estimates within the full LCR model, with the mean absolute difference between the estimated coefficient parameters being 0.005 across the 6 predictor variables and intercept. The clustering performance when compared with the full LCR model was similar, with an ARI value of 0.76.
The predictor variable selection sampler was run on dataset 1. The variables $x_5$ and $x_6$ are accurately identified as being uninformative, being excluded across all iterations of the sampler run. The comparatively weaker effect of the variable $x_4$ is also detected by the method, with a decreased posterior inclusion probability of 0.71, lower than variables $x_1, x_2, x_3$ which are included in almost all iterations. 
\begin{table}
\centering
\begin{tabular}{ccccccc}
\hline
Variable & 1 & 2 & 3 & 4 & 5 & 6    \\
\hline
True coefficient value & 1.20 & 1.00 & 0.80 & 0.40 & 0.00 & 0.00\\
PIP $(N = 150)$ & 0.93 & 0.80 & 0.29 & 0.14 & 0.03 & 0.01  \\
PIP $(N = 200)$ & 0.99 & 0.99 & 0.23 & 0.19 & 0.03 & 0.00  \\
PIP $(N = 300)$ & 1.00 & 1.00 & 0.49 & 0.28 & 0.00 & 0.00  \\
PIP $(N = 400)$ & 1.00 & 1.00 & 0.71 & 0.34 & 0.00 & 0.00  \\
PIP $(N = 500)$ & 1.00 & 1.00 & 0.99 & 0.71 & 0.00 & 0.00  \\

\hline
\end{tabular}
\caption{Table containing the true predictor coefficient values and the posterior inclusion probabilities (PIP) for each predictor variable in the LCR model with simultaneous selection for dataset 1, with varying sample sizes. For $N<500$, the PIP is computed as the mean PIP across 10 different random subsamples.}\label{table:predictor_selection_results_sim_study1}
\end{table}

The joint item-and-predictor variable selection yielded very similar posterior inclusion probabilities for predictors. The sample size sensitivity results of the predictor variable selection method for Simulation 1 are contained in Table \ref{table:predictor_selection_results_sim_study1}. Even for small sample sizes, Variables 1 and 2, which have the strongest effect on the model, are included with high probability.
Clustering performance was again similar to the other model variants when comparing with the true labels. The value of the ARI between the true labels and the fitted groups for the full variable selection is 0.76.
\subsection{Dataset 2}\label{subsec:sim_study2}
For the second dataset, we consider a 3 group scenario, to mimic a scenario of healthy, moderate and severe profiles. Table S4 contains the item probability parameters for dataset 2. We have $M = 13$ item variables with levels varying between 2 and 5. Variables 1--8 are discriminative, while 9--13 are not.
For dataset 2, covariate effects are varied in magnitude. We again consider $P = 6$ independent possible predictor variables, two of which are noise variables. See Table \ref{table:predictor_selection_results_sim_study2} for the relevant predictor coefficient values, with $\bsbet_3 = \mathbf{0}$ as the baseline.
\subsubsection{Group Identifiability Issues with the Full LCR Model}
When applying the full LCR model to dataset 2, an issue arises relating to group separation. This is likely due to the noise variables 9, 10, 11, 12, 13 obscuring some of the distinction between the groups. Hence when the sampler for the full LCR model is run, all observations are assigned to two of the three groups for almost all iterations. This prevents identification of the coefficient parameters $\bsbet_g$ due to the presence of two unconstrained coefficient parameters being estimated for two outcome variables with the absence of a baseline outcome. In order to make reliable inference, it is important to carry out item variable selection to remove the noise items that are causing this inaccuracy in estimation. The same issue was encountered when running predictor variable selection alone since all items were still being included. The parameter estimates for this model are therefore also unreliable. 
\subsubsection{Item Selection Models}
The model applying item selection alone accurately identifies the useful clustering variables 1, 2, 3, 4, 5, 6, 7, 8 with a posterior inclusion probability of 1 for each variable. Similarly, non-discriminating variables are successfully identified, with the posterior probability of inclusion not exceeding 0.01 for any of the variables 9, 10, 11, 12, 13. For posterior inclusion probabilities for each item variable, see Table S5. Estimates for the item probability parameters were computed post-hoc, and are accurate to a large degree when compared with the true simulated parameters. These values are provided in Table S4. We provide a summary of the $\bsbet$ estimation through ridgeline plots in Figure S2. For all of the coefficient parameters excluding those corresponding to the variable $X_2$, the 95\% HDI from the MCMC samples of the posterior contained the true parameter value.
The ARI value for this grouping when compared with the true labels was 0.67, which seems like a modest level of agreement, however the ARI value for clusters obtained using the true model parameters was 0.69 so this is a comparatively strong result. 
\begin{table}
\centering
\begin{tabular}{crrrrrr}
\hline
\multicolumn{1}{c}{Variable} & \multicolumn{1}{c}{1} & \multicolumn{1}{c}{2} & \multicolumn{1}{c}{3} & \multicolumn{1}{c}{4} & \multicolumn{1}{c}{5} & \multicolumn{1}{c}{6} \\
\hline
True coefficient value (Group 1) & 1.00 & $-1.00$ & 0.50 & $-0.40$ & 0.00 & 0.00 \\
True coefficient value (Group 2) & $-1.00$ & 1.00 & $-0.50$ & 0.40 & 0.00 & 0.00 \\
PIP $(N = 150)$ & 0.78 & 0.68 & 0.39 & 0.01 & 0.00 & 0.00 \\
PIP $(N = 200)$ & 0.85 & 0.93 & 0.48 & 0.01 & 0.00 & 0.00 \\
PIP $(N = 300)$ & 1.00 & 1.00 & 0.92 & 0.08 & 0.00 & 0.00 \\
PIP $(N = 400)$ & 1.00 & 1.00 & 1.00 & 0.21 & 0.00 & 0.00 \\
PIP $(N = 500)$ & 1.00 & 1.00 & 1.00 & 0.63 & 0.00 & 0.00 \\
\hline
\end{tabular}
\caption{Table containing the true predictor coefficient values and the posterior inclusion probabilities (PIP) for each predictor variable in the LCR model with simultaneous selection for dataset 2, with varying sample sizes. For $n<500$, the PIP is computed as the mean PIP across 10 different random subsamples.}\label{table:predictor_selection_results_sim_study2}
\end{table}
The sampler for the simultaneous item variable and predictor variable selection was run. The item variables 9, 10, 11, 12, 13 were again correctly identified as being non-discriminative for grouping, with posterior inclusion probabilities comparable with the values found when performing item selection alone, and the informative variables 1, 2, 3, 4, 5, 6, 7, 8 each had a posterior inclusion probability of 1 (see Table S5). For predictor variable selection, the variables $x_1$, $x_2$ and $x_3$ all again had posterior inclusion probabilities of 1, while the corresponding values for variables $x_4$ was 0.63, and the probabilities for inclusion of the noise variables $x_5$ and $x_6$ were both 0 (see Table \ref{table:predictor_selection_results_sim_study2}). These values aligned with the magnitudes of the true coefficient values. Results from the sample size sensitivity assessment for dataset 2 are contained in Table \ref{table:predictor_selection_results_sim_study2}. As with the first simulation study, variables with the strongest effect on the model are included with reasonable probability. 
As in the item selection model the value of the ARI between the true labels and the groups fitted by the variable selection model was high when compared with the ARI for the true parameter values, with a value of 0.67.
\subsection{Sample Size Calibration}\label{subsec:samplesize_calibration}
To facilitate interpretation in applications of similar scale, the value of $N = 150$ from each of the simulation studies (see Table \ref{table:predictor_selection_results_sim_study1}, \ref{table:predictor_selection_results_sim_study2}) serves as a sample size calibration target for the predictor variable selection models. At $N \approx 150$, truly large effects ($|\bsbet| \approx 1.0$ to $1.2$) yield high PIP values (PIP $\approx 0.7$ to $0.9$), while moderate and weak effects are markedly lower. This gives us a practical reference for datasets of comparable size, such as the CSHQ data with $N = 148$.
\section{Data Analysis}\label{sec:results}
We apply the sampling method to the CSHQ sleep patterns data set described in Subsection \ref{subsec:sleep_data}. 
\subsection{Model Selection}
The LCA collapsed sampler of \citet{white_bayesian_2016} was run for the CSHQ sleep patterns dataset as the initial step of the model selection procedure. From the output of this model, the clustering point estimation method of \citet{wade_bayesian_2018} described in \ref{subsubsec:clust_point_est} was applied. The resulting clustering point estimate consisted of four groups of varying size. This was a surprising result, since the collapsed sampler gave the highest posterior probability to 6 groups, with a small probability for 5, 7 and 8 groups. This is likely due to the presence of a small number of unusual observations within the dataset, rather than meaningful subgroups. The smallest three groups of this 6 group solution also had very small proportions of 0.06, 0.03 and 0.03 respectively, which suggests that a 6 group solution may be overfit and would not generalize well to larger sample sizes. Hence it was of interest to assess the plausibility of a 4 group LCA model for the data. In order to do this, the 95\% credible ball was computed for the clustering point estimate derived from the collapsed sampler, as outlined in \ref{subsubsec:clust_point_est}. The computed value for the VI distance from the point estimate to the horizontal bound of the credible ball was 1.44. A 4 group LCA model was fit and observations were assigned to groups, again by minimising the posterior expected variation of information. When running this LCA, simultaneous item variable selection was carried out using the collapsed sampler in order to keep consistency in the configuration of variables. The VI distance from the point estimate of the collapsed sampler to the 4 group point estimate was 0.52. Since this value is smaller than the distance to the horizontal bound, this LCA solution is contained within the 95\% credible ball and hence is a plausible model to represent the data. Along with this, there was a strong level of agreement between the clustering point estimate from the collapsed sampler and that of the 4 group LCA model, with an ARI value of 0.86.  For these reasons, the value of $G$ was chosen to be 4 for further analysis. 
For the analysis that follows, the 4 groups are denoted as Groups A, B, C and D. We denote the predictor variables of age, gender, ASD diagnosis, Intellectual Disability diagnosis and other comorbidities as $x_1, x_2, x_3, x_4, x_5$ with corresponding coefficient parameters $\bsbet_{g1}, \bsbet_{g2}, \bsbet_{g3}, \bsbet_{g4}, \bsbet_{g5}$ respectively. For interpretability, for this dataset we consider $g = \text{B},\text{C},\text{D}$ ($\bsbet_{\text{A}} = \mathbf{0}$ as the baseline group, since it is the largest group).
\subsection{Group Characteristics}
\begin{figure}
    \centering
    \includegraphics[width=\linewidth]{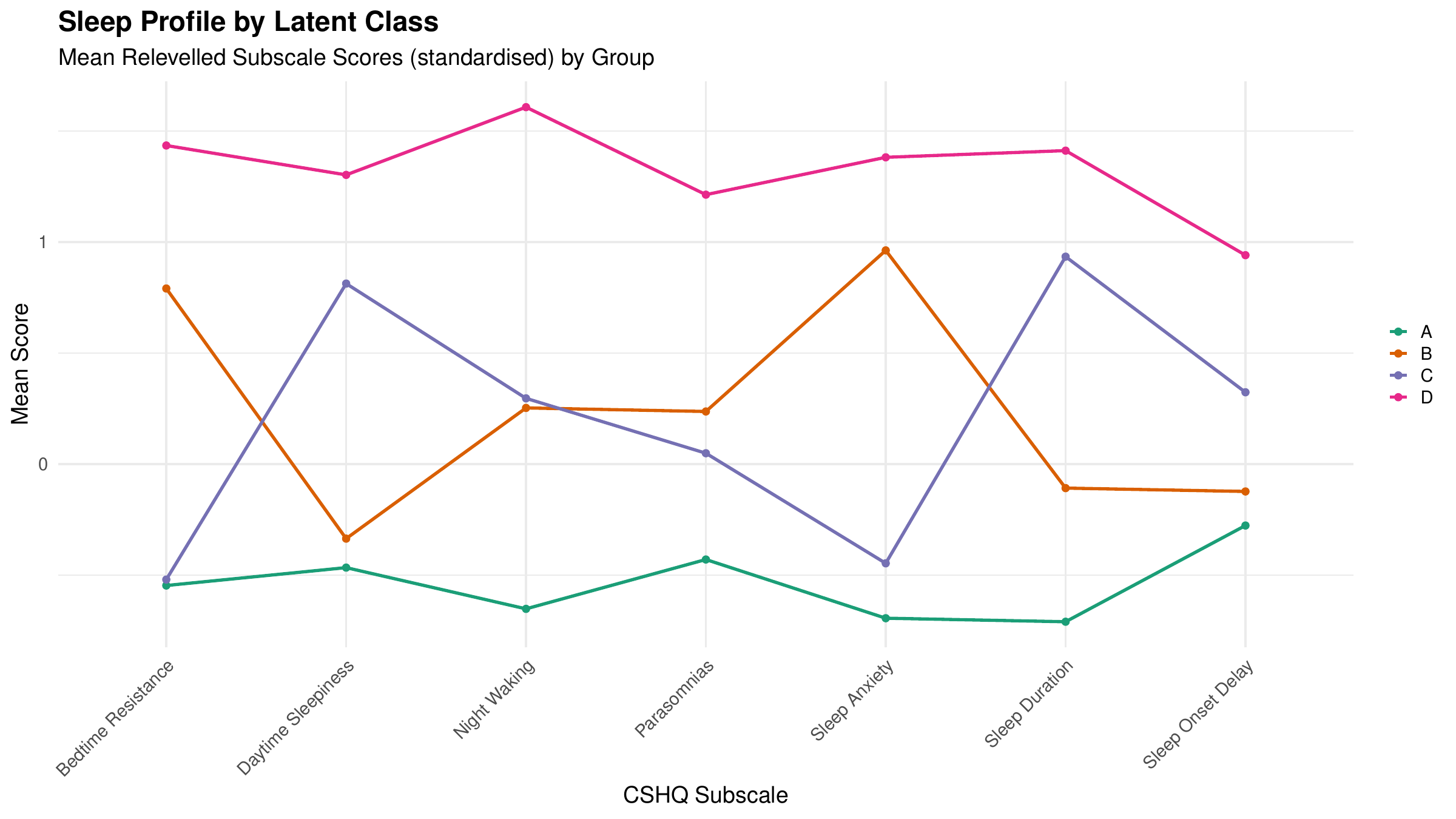}
    \caption{Plot of sleep profiles by group. For each individual, subscale totals were re-levelled by subtracting the minimum possible score for that subscale. The plotted values are the mean of these re-levelled subscale scores across individuals assigned to each LCA group, following standardisation.}
    \label{fig:CSHQ_profile_plot}
\end{figure}

Item probability estimates were computed post-hoc from the initial 4 group LCA run with variable selection. The item variables that had a posterior inclusion probability of more than 0.5 were considered to be informative for distinguishing groups, and a full set of mosaic plots for these variables are provided in Figures S3 and S4. Figure \ref{fig:CSHQ_profile_plot} contains a visualisation of the varying patterns of sleep disturbances between groups. This summarises the survey responses of each group, aggregated within each of the eight CSHQ subscales. Subscale totals were re-levelled by subtracting the minimum possible score for each subscale, with mean values of the re-levelled subscale scores for individuals assigned to each LCA group plotted, following standardisation. Lower values are hence indicative of reduced sleep difficulties. 

Clear differences in variable response patterns across groups can be observed. Notably, children in Group A and Group D record the lowest and highest degree of sleep difficulties across all subscales, while Groups B and C exhibit contrasting challenges.  

In the following paragraphs, we describe the characteristic sleep patterns for each latent class in more detail. Item codes (e.g., SD2, BR5) refer to specific CSHQ questions, with accompanying values indicating the posterior mean probability of endorsing the 'sometimes' or 'usually' response category for that item within the class. Table~\ref{table:groups_summary_table} contains summary information for each of the 4 groups, where the values of mean CSHQ total score, mean age and sleep disorder proportion are computed across the observations assigned using the clustering point estimate. 
\begin{table}%[H]
\begin{center}
\begin{tabular}{ccccc}
    \hline
    $g$ & $\pi_g$ (sd) & Mean Total CSHQ (sd) & Mean Age (sd) & Sleep Disorder Proportion \\
    \hline
    A & 0.42 (0.05) & 39.90 (3.92) & 7.44 (3.28) &  0.41 \\
    B & 0.28 (0.04) & 49.91 (4.79) & 6.29 (2.10) &  1.00 \\
    C & 0.21 (0.04) & 50.45 (5.17) & 8.27 (3.50) &  1.00 \\
    D & 0.09 (0.03) & 66.92 (3.15) & 7.50 (2.68) &  1.00 \\
    \hline                                            
\end{tabular}            
\caption{Summary of the 4 groups identified by LCA. Sleep disorder proportion refers the number of participants in a group above the threshold indicating a sleep disorder in the CSHQ survey (>41)}\label{table:groups_summary_table}
\end{center}
\end{table}
\subsubsection{Group A: Restful/Healthy Sleepers}
Group A represented the largest proportion of the sample and was characterized by a generally healthy sleep profile across the various CSHQ subscales. The mean total CSHQ score of 39.90 for this group was below the established threshold for clinically significant sleep problems (<41), indicating that, as a whole, children in this group did not exhibit patterns suggestive of a sleep disorder. 

Despite this overall favourable profile, a closer examination of individual item probabilities reveals that children in Group A exhibited slightly increased probabilities of higher scores in the subscales of Sleep Onset Delay, Sleep Anxiety, and Daytime Sleepiness. This might suggest that, while the majority of sleep behaviors in this group were comparatively healthy, there was a mild tendency toward occasional difficulties in initiating sleep (SOD1: 0.41), some bedtime-related anxiety like being afraid to sleep in the dark (SA2: 0.29) and having trouble sleeping away (SA4: 0.29), and some problems with daytime tiredness, particularly in their difficulty with getting out of bed (DS4: 0.38) , frequency in presenting as tired (DS6: 0.35) and waking in a negative mood (DS2: 0.30). Along with this, there were issues prevalent with Parasomnias, particularly with sleep talking (P2: 0.31) and restlessness and movement (P3: 0.52), although the latter appeared to be a prominent issue across all of the groups. However, these elevations were overall modest and did not approach the levels observed in some of the other groups.
\subsubsection{Group B: Anxious Nighttime Sleepers} 
Group B, the second largest subgroup identified, is distinguished by a sleep profile marked by elevated anxiety and regulatory difficulties at night. Children in this group have a strong tendency toward bedtime resistance, with frequent struggles at bedtime (BR5: 0.90) and persistent difficulties in settling down to sleep. Children in this group show consistently higher probabilities of scoring in the upper categories for items related to sleep anxiety, such as reluctance to sleep in the dark (SA2: 0.67), difficulty sleeping alone (SA3: 0.93), and a strong need for parental presence at bedtime (SA1: 0.86).

In addition to anxiety-driven behaviors, this group is characterized by significant problems with sleep initiation, as evidenced by a high probability of taking more than 20 minutes to fall asleep (SOD1: 0.58). Night waking is also prevalent, with children in this group commonly awakening at least once during the night (NW2: 0.75) further disrupting their sleep continuity. Children in this group are also likely to often move into others beds during the night (NW1: 0.59). Restlessness during sleep is a notable feature, with frequent movement and difficulty maintaining restful sleep throughout the night (P3: 0.74). Along with this, sleep talking is also common among children in this group (P2: 0.56).

 The consequences from these difficulties are apparent in the Daytime Sleepiness subscale, as children in Group B are more likely to present as tired during the day (DS6: 0.53), reflecting the impact of fragmented and insufficient sleep. The combination of heightened bedtime anxiety, persistent resistance, frequent night waking, and daytime sleepiness suggests that this group experiences a varied pattern of sleep disturbance, with anxiety and difficulties in sleep regulation as central features.
\subsubsection{Group C: Short/light Sleepers} 
Group C, comparable in size to Group B, presents a distinct sleep profile despite exhibiting a similar magnitude of mean total CSHQ scores. While the global sleep disturbance levels appear close between these groups, a detailed examination of item-level probabilities reveals fundamentally different patterns of sleep difficulty. Group C is characterized primarily by problems with sleep quantity and timing rather than the anxiety-driven difficulties observed in Group B.

The primary distinguishing feature of this group centres on sleep duration inadequacies, with children having a high probability of demonstrating a consistent pattern of insufficient sleep across multiple indicators. Children in this group exhibit elevated probabilities of sleeping too little on a nightly basis (SD1: 0.87), failing to maintain appropriate sleep amounts (SD2: 0.83), and struggling with day-to-day consistency in their sleep duration (SD3: 0.79).

Along with these duration difficulties, Group C children are also more likely to show problems with sleep initiation, with many requiring more than 20 minutes to fall asleep (SOD1: 0.82). This combination of delayed sleep onset and shortened sleep duration creates a particularly problematic pattern where children experience both difficulties accessing sleep and insufficient total sleep time. This is compounded with a restlessness (P3: 0.75) which likely impacts ability to obtain sufficient sleep quantities.

Again, the consequences of this sleep profile manifest prominently during daytime hours. Children in Group C are characterized by significant morning difficulties, including waking up in negative moods (DS2: 0.81) and experiencing considerable challenges getting out of bed (DS4: 0.70). Many of these children frequently present as visibly tired (DS6: 0.85), possibly reflecting the cumulative impact of consistent sleep insufficiency. This daytime symptom pattern suggests that Group C represents children whose sleep difficulties primarily stem from quantitative rather than qualitative sleep problems, distinguishing them from the anxiety-based sleep disruptions characteristic of Group B.  
\subsubsection{Group D: Pervasive Sleep Difficulties}
Group D, although the smallest subgroup, represents children with the most severe and wide ranging sleep difficulties. Unlike the more narrowly defined pattens seen in Groups B and C, the sleep profile of this group is characterised by consistently high probabilities of disturbed sleep across nearly all domains. Children in Group D have a very high probability of inadequate sleep duration, with probabilities above 0.90 for regularly sleeping too little (SD1: 0.91), and failing to achieve sufficient overall sleep amounts (SD2: 0.93). Bedtime and night regulation difficulties are also more pronounced, with elevated probabilities of struggling at  bedtime (BR5: 0.89), needing parental presence to fall asleep (SA1: 0.88), difficulty sleeping alone (SA3: 0.89) and frequent night waking (NW2: 0.91). Parasomnias are also prevalent, particularly restlessness during sleep (P3: 0.86), alongside heightened rates of sleep anxiety difficulties such as fears of sleeping in the dark (SA2: 0.68) and challenges with sleeping away from home (SA4: 0.87).

Similarly to Groups B and C, the effects of these nighttime difficulties are evident in daytime functioning. Children in Group D are highly likely to present as tired (DS6: 0.92), struggle to get out of bed in the morning (DS4: 0.76), and frequently appear in a negative mood on waking (DS2: 0.81). These daytime impairments reflect the cumulative burden of both fragmented sleep and insufficient sleep quantity, differentiating this group not by a single core difficulty but by the range and severity of problems across all sleep quality subscales. When considered together, this profile indicates that Group D captures a subgroup of children with global sleep dysregulation, in which multiple sleep processes are simultaneously impared to a high degree. 
\subsection{LCR Model}
\subsubsection{Model Identifiability Issues with the Full LCR Model}
Initially, the full LCR model that excluded item variable selection was run on the data set. This full model of 33 items and 5 predictors exhibited signs of severe non-identifiability, which was indicated by the failure of convergence to a stable posterior distribution observed in the trace plots of the log posterior. 
This is most likely caused by a model which is over-parameterised due to the inclusion of all variables, making it impossible to obtain meaningful estimates for parameters without the use of a more parsimonious specification. Similarly, when predictor variable selection was run on its own, similar phenomena were observed in the trace plots. Hence, similar to Simulation 2, it is important to apply item selection in order to obtain reliable parameter estimates.
\subsubsection{Item Selection Models}\label{subsec:CSHQ_item_sel}
\begin{figure}
    \centering
    \includegraphics[width=\linewidth]{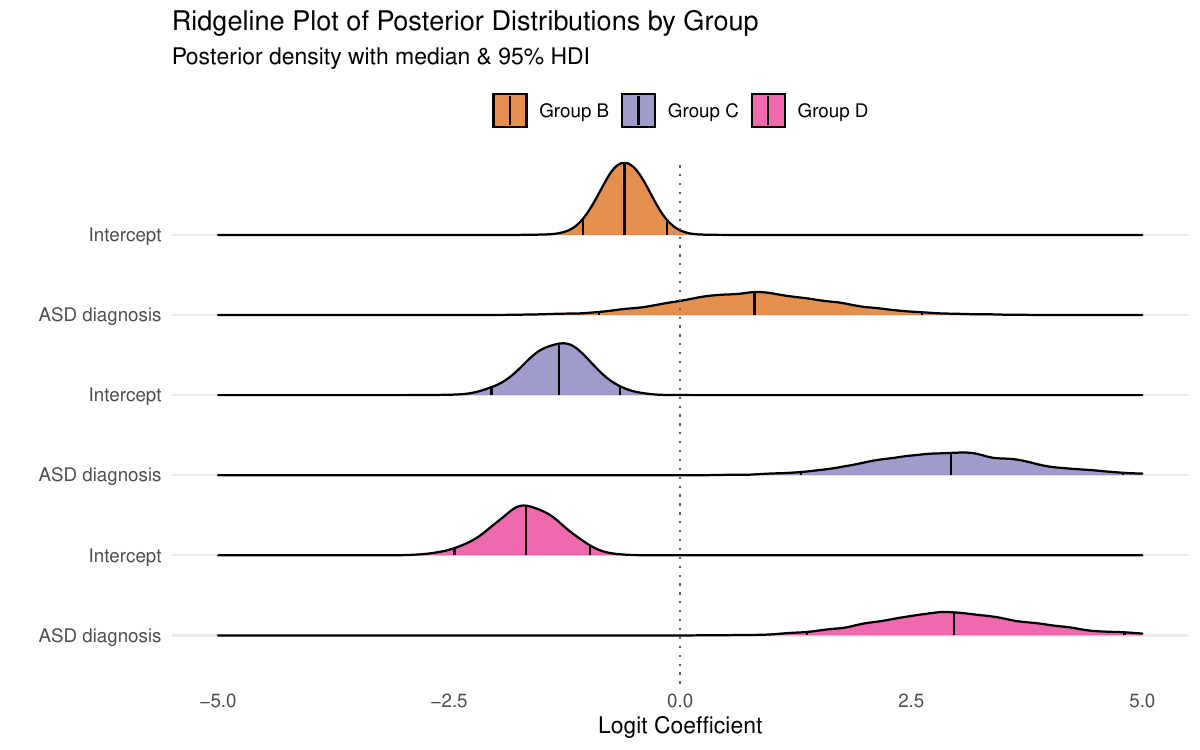}
    \caption{Ridgeline plots of posterior distributions for the intercept and ASD diagnosis coefficients in the variable selection model for the CSHQ sleep patterns data.}
    \label{fig:CSHQ_ridgeline_combined_ASD_only}
\end{figure}

The item selection model included 21 of the variables more than half of the time, all of which were consistent with the results from the preliminary run of the collapsed sampler. This is to be expected since a variable that is sufficiently discriminative within the LCA model would also be discriminative to a similar degree in the regression model. These retained items were bedtime resistance 2, 3, 5, sleep onset delay, sleep duration 1, 2, 3, sleep anxiety 1, 2, 3, 4, night waking 1, 2, 3, parasomnias 2, 3, 5, and daytime sleepiness 2, 4, 5, 6. The other 12 variables were less discriminative, being included less than 0.5 of the time. Table S10 includes the posterior probability of inclusion of each variable. 
During estimation, we encountered separation for the intellectual-disability and other-diagnosis indicators in the multinomial logit because Group B had (near) zero cases with these diagnoses. The model thus tended to perfectly predict non-membership in Group B, yielding unstable, sparsity-driven estimates. We therefore imposed a narrower prior on these coefficients with variance $5^2$ for regularisarion, constraining estimates to a plausible range, and preventing divergence. This pattern may still reflect a clinically relevant feature of the `anxious' group. A summary of the estimation of the $\bsbet$ coefficient parameters without considering predictor variable selection is provided in Table S8. 
The model with predictor variable selection was run using the same variables as previously for candidate predictors. Table \ref{table:predictor_selection_results_CSHQ} contains the relevant posterior inclusion probability estimates from this sampler. Interestingly, the variable of age was included in none of the iterations of the sampler run, despite the 95\% HDI for Group B excluding zero in the model without predictor variable selection. This may be due to the fact that if age is discriminative for membership in only one out of the three non-baseline groups then it would make it less likely to warrant inclusion in the model on a given iteration. The variables of ID diagnosis and other diagnoses were included in a marginal proportion of iterations, while the variable of ASD diagnosis had an elevated posterior inclusion probability of 0.70, indicating that ASD diagnosis was comparatively informative for predicting group membership. This result aligns with the posterior distributions of the $\beta$ coefficients in the full item selection model, where the coefficients for this variable for groups C and D were found to be credibly positive. Figure \ref{fig:CSHQ_ridgeline_combined_ASD_only} contains ridgeline plots for the intercept and ASD diagnosis regression coefficients for the simultaneous variable selection model. These are created by only including the iterations where the ASD diagnosis variable was included in the model.   
\begin{table}
\centering
\begin{tabular}{cccccc}
\hline
Variable & Age & Gender & ASD Diagnosis & ID Diagnosis & other diagnoses    \\
\hline
PIP  
& 0.00 & 0.00 & 0.70 & 0.02 & 0.00  \\
\hline
\end{tabular}
\caption{Table containing the posterior inclusion probabilities (PIP) for each predictor variable in the LCR model with simultaneous variable selection for the CSHQ dataset.}\label{table:predictor_selection_results_CSHQ}
\end{table}
\section{Discussion}\label{sec:discussion}
In this paper, we introduced and evaluated a fully Bayesian framework for inference in the Latent Class Regression (LCR) model, incorporating simultaneous variable selection for both predictors and item responses. The key methodological contribution lies in adapting the Pólya-Gamma augmentation to the LCR setting, enabling Gibbs sampling of regression coefficients. By extending to include variable selection mechanisms through predictor inclusion indicators and partial collapsing for item variables, we provided a flexible framework that improves inference and enhances interpretability by automatically identifying the most relevant features for subgroup differentiation.

Our simulation studies demonstrated that, while the LCR model can suffer from weak identifiability and poor mixing in the presence of noise variables, the addition of item variable selection successfully mitigates these issues. Removing uninformative items clarified latent subgroup structures and yielded stable parameter estimates. Similarly, predictor variable selection highlighted the covariates most strongly associated with group membership, with posterior inclusion probabilities providing a natural measure of uncertainty. 

The application to the Children's Sleep Habits Questionnaire (CSHQ) further underscored the utility of performing item selection, where identifiability was again recovered after applying item variable selection. The traditional cumulative scoring system of the CSHQ obscures heterogeneity  in children's sleep difficulties, whereas the approach detailed here identified meaningful subgroups with distinct behavioural profiles. These groups --- anxious nighttime sleepers (Group B), short/light sleepers (Group C), pervasive sleep difficulties (Group D) --- suggest a need for tailored intervention strategies. For instance, interventions focusing on anxiety management may be more beneficial for participants from Group B, while participants within Group C may respond better to strategies targeting sleep scheduling or duration. The strong predictive influence of Autism Spectrum Disorder (ASD) diagnosis also highlights the importance of considering neurodevelopmental conditions when modelling paediatric sleep difficulties, due to the elevated probabilities of membership in groups with more prevalent sleep difficulties for participants with ASD. In particular, ASD diagnosis positively predicted membership of the short/light sleepers and the pervasive sleep difficulties profiles, rather than all problematic sleep groups, indicating that ASD is linked to specific patterns of sleep disruption rather than a general increase in sleep problems. 

Several limitations should be acknowledged. First, the current framework requires the number of groups $G$ to be estimated in a separate preliminary step. This is justifiable from a modelling perspective, however it externalises uncertainty in $G$ and prevents fully joint inference. Second, quasi-separation effects from more uncommon diagnoses led to unstable regression coefficient estimates, indicating the need for regularisation strategies without a larger, more balanced sample. Third, while the current model treats the CSHQ responses as nominal categorical variables, when they are actually ordinal responses, where the distance between the responses `rarely', `sometimes' and `usually' may carry additional information. Treating these ordinal responses as nominal may obscure subtleties in the gradations of severity or frequency of variables, which points to a useful avenue for methodological extension.

Future work should address these limitations in several ways. Integrating model-based selection of the number of groups $G$ directly into the Bayesian LCR framework would eliminate the reliance on preliminary estimation and allow full quantification of the uncertainty for the value of $G$. Extending the methodology to handle ordinal responses directly would more faithfully capture the structure of the sleep questionnaire data and could yield stronger insights into subgroup properties. Another promising direction is methodological refinement through the Ultimate Pólya-Gamma sampler of \citet{zens_ultimate_2024}, which has been shown to improve efficiency and mixing under severe imbalance and high dimensionality. Incorporating this into the LCR framework could improve scalability of the method. Finally, further exploration of robustness methods for handling outliers or noisy observations would help prevent the introduction of extraneous groups within the model, ensuring that subgroup solutions remain clinically meaningful. 
\paragraph{Acknowledgements.}
These data were collected by Lynn Walsh in partial fulfilment of the MSc Applied Behaviour Analysis at the School of Psychology, Trinity College Dublin.

\paragraph{Funding.}
The work of all authors was supported by a Trinity Research Doctorate Award, an award from Trinity College Dublin. JW's work on this manuscript was also supported by Taighde Éireann – Research Ireland under Grant number 21/FFP-P/10123.  

\section*{Supplementary Material}

\noindent\textbf{Additional Results:} We provide additional results, tables and visualisations from simulation studies and data analysis, as well as technical derivations to supplement the above results.

\bibliographystyle{apalike-doi} 
\bibliography{references}      

\end{document}

\maketitle
\tableofcontents

\section{Simulation Studies}
\subsection{Simulation 1: Parameter Estimates and Selection Results}

\begin{figure}[H]
    \centering
    \includegraphics[width=\linewidth]{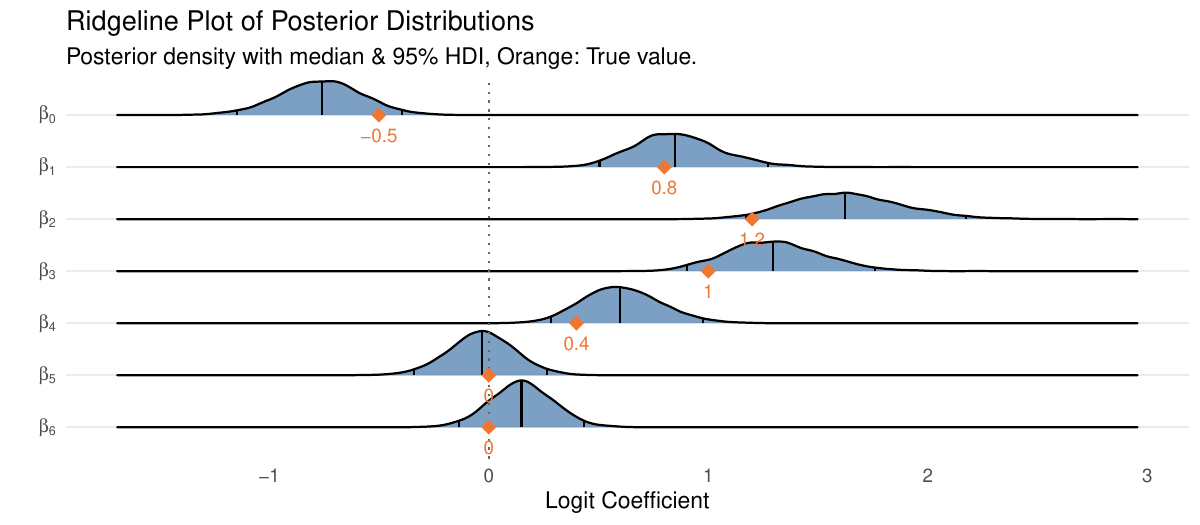}
    \caption{Ridgeline plots of posterior distribution of regression coefficients for simulation 1.}
    \label{fig:sim1_ridgeline_plot}
\end{figure}

Table \ref{table:sim1_item_prob_estimates_full_LCR} contains the true and estimated values of the item probability parameters for the full LCR model for simulation 1, computed by posterior mean from Gibbs sampling. Table \ref{table:sim1_item_prob_estimates_itemsel} contains the estimates for the item probability parameters from the item selection model computed post-hoc. Table \ref{table:item_selection_results_sim_study1} contains posterior inclusion probabilities for each item variable in simulation 1. Figure \ref{fig:sim1_ridgeline_plot} contains a ridgeline plot illustrating the posterior distributions of the regression coefficients for the full LCR model without variable selection applied to simulation 1.

\begin{table}[H]
  \centering
\begin{tabular}{llcccc}
\toprule
\textbf{Var.} & \textbf{Cat.} & \textbf{True 1} & \textbf{Est. 1 (sd)} & \textbf{True 2} & \textbf{Est. 2 (sd)}  \\
\midrule
1 & 1 & 0.15 & 0.08 (0.03) & 0.70 & 0.74 (0.03)    \\
  & 2 & 0.25 & 0.30 (0.03) & 0.20 & 0.16 (0.02)   \\
  & 3 & 0.60 & 0.61 (0.04) & 0.10 & 0.10 (0.02)   \\
2 & 1 & 0.20 & 0.23 (0.03) & 0.55 & 0.55 (0.03)   \\
  & 2 & 0.35 & 0.32 (0.03) & 0.30 & 0.29 (0.03)   \\
  & 3 & 0.45 & 0.45 (0.04) & 0.15 & 0.15 (0.02)   \\
3 & 1 & 0.10 & 0.14 (0.03) & 0.80 & 0.82 (0.03)    \\
  & 2 & 0.15 & 0.15 (0.03) & 0.15 & 0.13 (0.02)    \\
  & 3 & 0.75 & 0.71 (0.04) & 0.05 & 0.05 (0.02)    \\
4 & 1 & 0.25 & 0.26 (0.03) & 0.45 & 0.44 (0.03)    \\
  & 2 & 0.40 & 0.35 (0.04) & 0.35 & 0.37 (0.03)    \\
  & 3 & 0.35 & 0.40 (0.04) & 0.20 & 0.19 (0.02)    \\
5 & 1 & 0.40 & 0.39 (0.04) & 0.40 & 0.44 (0.03)   \\
  & 2 & 0.50 & 0.53 (0.04) & 0.50 & 0.45 (0.03)   \\
  & 3 & 0.10 & 0.08 (0.02) & 0.10 & 0.11 (0.02)   \\
6 & 1 & 0.70 & 0.75 (0.03) & 0.70 & 0.72 (0.03)    \\
  & 2 & 0.10 & 0.07 (0.02) & 0.10 & 0.09 (0.02)   \\
  & 3 & 0.20 & 0.18 (0.03) & 0.20 & 0.19 (0.02)   \\
7 & 1 & 0.10 & 0.08 (0.02) & 0.10 & 0.09 (0.02)   \\
  & 2 & 0.50 & 0.50 (0.04) & 0.50 & 0.52 (0.03)    \\
  & 3 & 0.40 & 0.42 (0.04) & 0.40 & 0.38 (0.03)    \\  
8 & 1 & 0.33 & 0.28 (0.03) & 0.33 & 0.32 (0.03)  \\
  & 2 & 0.33 & 0.34 (0.03) & 0.33 & 0.34 (0.03)    \\
  & 3 & 0.33 & 0.38 (0.04) & 0.33 & 0.34 (0.03)   \\  
\bottomrule
\end{tabular}
\caption{True and estimated values of the item probability parameters for the full LCR model for simulation 1.}
  \label{table:sim1_item_prob_estimates_full_LCR}
\end{table}

\begin{table}[H]
  \centering
\begin{tabular}{llcccc}
\toprule
\textbf{Var.} & \textbf{Cat.} & \textbf{True 1} & \textbf{Est. 1 (sd)} & \textbf{True 2} & \textbf{Est. 2 (sd)}  \\
\midrule
1 & 1 & 0.15 & 0.08 (0.03) & 0.70 &  0.73 (0.03)   \\
  & 2 & 0.25 & 0.30 (0.03) & 0.20 &  0.16 (0.02)  \\
  & 3 & 0.60 & 0.62 (0.04) & 0.10 &  0.11 (0.02)  \\
2 & 1 & 0.20 & 0.23 (0.03) & 0.55 &  0.55 (0.03)   \\
  & 2 & 0.35 & 0.32 (0.03) & 0.30 &  0.29 (0.03)   \\
  & 3 & 0.45 & 0.45 (0.04) & 0.15 &  0.15 (0.02)  \\
3 & 1 & 0.10 & 0.13 (0.03) & 0.80 &  0.82 (0.03)  \\
  & 2 & 0.15 & 0.15 (0.03) & 0.15 &  0.13 (0.02)    \\
  & 3 & 0.75 & 0.72 (0.04) & 0.05 &  0.05 (0.02)  \\       
4 & 1 & 0.25 & 0.26 (0.03) & 0.45 &  0.44 (0.03)   \\
  & 2 & 0.40 & 0.35 (0.04) & 0.35 &  0.37 (0.03)   \\
  & 3 & 0.35 & 0.40 (0.04) & 0.20 &  0.19 (0.02)   \\
5 & 1 & 0.40 & 0.40 (0.04) & 0.40 &  0.44 (0.03) \\
  & 2 & 0.50 & 0.52 (0.04) & 0.50 &  0.46 (0.03)  \\
  & 3 & 0.10 & 0.08 (0.02) & 0.10 &  0.11 (0.02)  \\
6 & 1 & 0.70 & 0.74 (0.04) & 0.70 &  0.72 (0.03)   \\
  & 2 & 0.10 & 0.07 (0.02) & 0.10 &  0.09 (0.02)  \\
  & 3 & 0.20 & 0.18 (0.03) & 0.20 &  0.19 (0.02) \\
7 & 1 & 0.10 & 0.08 (0.02) & 0.10 &  0.09 (0.02)  \\
  & 2 & 0.50 & 0.50 (0.04) & 0.50 &  0.52 (0.03)   \\
  & 3 & 0.40 & 0.42 (0.04) & 0.40 &  0.39 (0.03)  \\  
8 & 1 & 0.33 & 0.29 (0.03) & 0.33 &  0.32 (0.03) \\
  & 2 & 0.33 & 0.34 (0.04) & 0.33 &  0.34 (0.03)  \\
  & 3 & 0.33 & 0.34 (0.04) & 0.33 &  0.34 (0.03)  \\  
\bottomrule
\end{tabular}
\caption{Estimated values of the item probability parameters for the item selection model for simulation 1.}
  \label{table:sim1_item_prob_estimates_itemsel}
\end{table}

\begin{table}[H]
\centering
\begin{tabular}{ccccccccc}
\hline
Variable & 1 & 2 & 3 & 4 & 5 & 6 & 7 & 8 \\
\hline
PIP (item) & 1.00 & 1.00 & 1.00 & 1.00 & 0.08 & 0.02 & 0.03 & 0.04  \\
PIP (simul) & 1.00 & 1.00 & 1.00 & 1.00 & 0.07 & 0.02 & 0.03 & 0.04  \\
\hline
\end{tabular}
\caption{Posterior inclusion probabilities (PIP) for each item variable in the LCR model for simulation 1. 
PIP (item): item selection alone; PIP (simul): simultaneous selection. }\label{table:item_selection_results_sim_study1}
\end{table}

\subsection{Simulation 2: Parameter Estimates and Selection Results}
\begin{figure}[H]
    \centering
    \includegraphics[width=\linewidth]{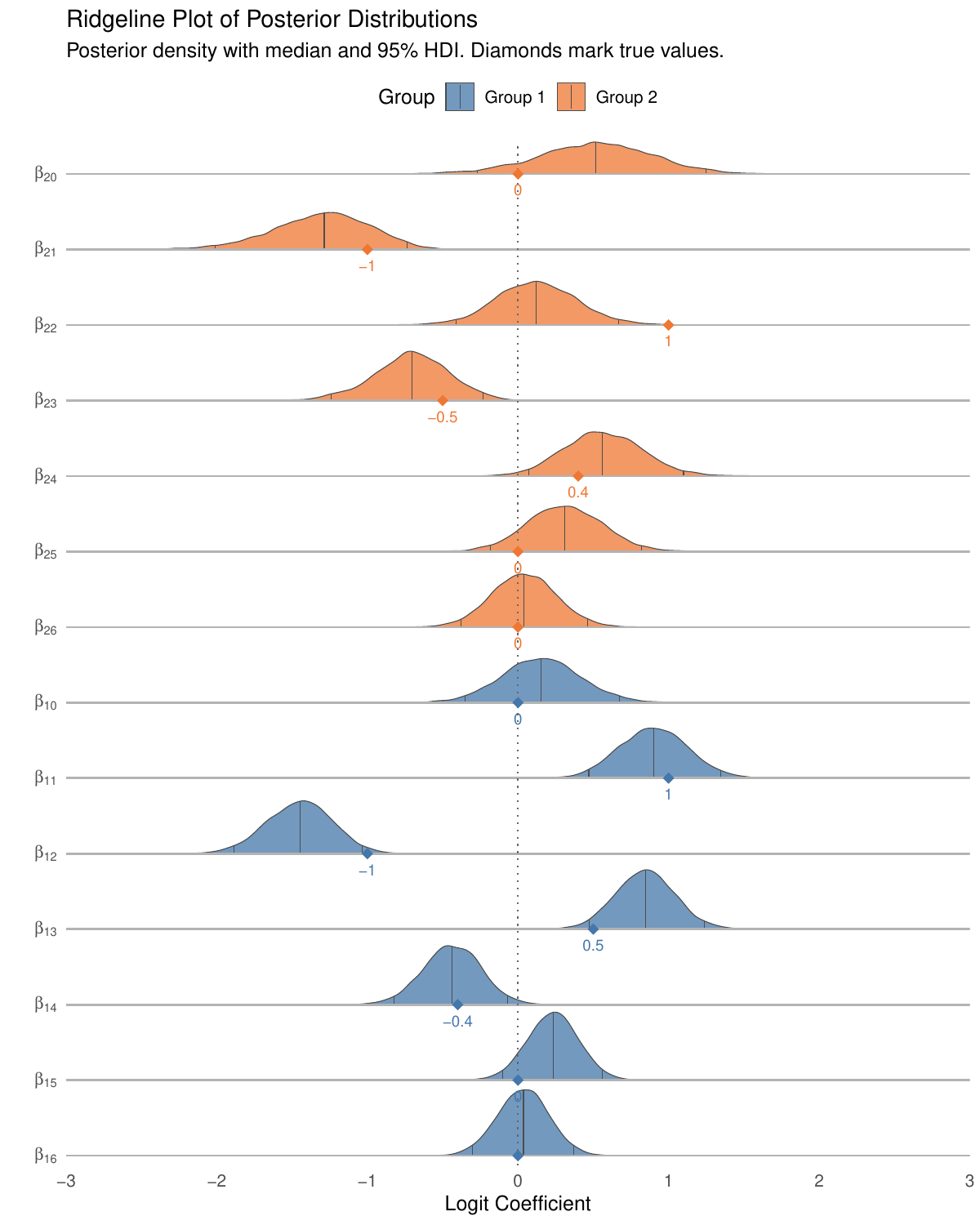}
    \caption{Ridgeline plots of posterior distribution of regression coefficients for simulation 2.}
    \label{fig:sim2_ridgeline_plot}
\end{figure}

Table \ref{table:sim2_item_prob_estimates_itemsel} contains the true and estimated values of the item probability parameters for the collapsed LCR model for simulation 2, computed post-hoc. Table \ref{table:item_selection_results_sim_study2} contains posterior inclusion probabilities for each item variable in simulation 2. Figure \ref{fig:sim2_ridgeline_plot} contains a ridgeline plot illustrating the posterior distributions of the regression coefficients for the LCR model with item selection only applied to simulation 2.
\begin{table}[H]\centering
\begin{tabular}{llcccccc}
\toprule
\textbf{Var.} & \textbf{Cat.} & \textbf{True 1} & \textbf{Est. 1 (sd)} & \textbf{True 2} & \textbf{Est. 2 (sd)} & \textbf{True 3} & \textbf{Est. 3 (sd)} \\
\midrule
1 & 1 & 0.15 & 0.18 (0.03) & 0.60 & 0.58 (0.04) & 0.80 & 0.80 (0.05)   \\
  & 2 & 0.85 & 0.82 (0.03) & 0.40 & 0.42 (0.04) & 0.20 & 0.20 (0.05)   \\
2 & 1 & 0.25 & 0.24 (0.03) & 0.45 & 0.45 (0.04) & 0.70 & 0.77 (0.06)  \\
  & 2 & 0.75 & 0.76 (0.03) & 0.55 & 0.55 (0.04) & 0.30 & 0.23 (0.06)   \\
3 & 1 & 0.70 & 0.70 (0.04) & 0.20 & 0.23 (0.03) & 0.65 & 0.71 (0.08)   \\
  & 2 & 0.30 & 0.30 (0.04) & 0.80 & 0.77 (0.03) & 0.35 & 0.29 (0.08)   \\
4 & 1 & 0.10 & 0.08 (0.02) & 0.35 & 0.39 (0.04) & 0.70 & 0.64 (0.07)   \\
  & 2 & 0.25 & 0.28 (0.03) & 0.40 & 0.34 (0.04) & 0.20 & 0.26 (0.06)   \\
  & 3 & 0.65 & 0.64 (0.04) & 0.25 & 0.27 (0.03) & 0.10 & 0.09 (0.04)   \\
5 & 1 & 0.20 & 0.18 (0.03) & 0.25 & 0.29  (0.04) & 0.65 & 0.66 (0.07)  \\
  & 2 & 0.15 & 0.17 (0.03) & 0.60 & 0.59  (0.04) & 0.25 & 0.25 (0.06)  \\
  & 3 & 0.65 & 0.65 (0.04) & 0.15 & 0.12  (0.03) & 0.10 & 0.09 (0.04)  \\
6 & 1 & 0.15 & 0.11 (0.03) & 0.50 & 0.45  (0.03) & 0.75 & 0.79 (0.06)   \\
  & 2 & 0.20 & 0.21 (0.03) & 0.35 & 0.38  (0.04) & 0.15 & 0.13 (0.05)  \\
  & 3 & 0.65 & 0.68 (0.04) & 0.15 & 0.16  (0.03) & 0.10 & 0.09 (0.03)  \\
7 & 1 & 0.10 & 0.10 (0.02) & 0.25 & 0.25  (0.03) & 0.60 & 0.51 (0.06)  \\
  & 2 & 0.15 & 0.13 (0.03) & 0.35 & 0.33  (0.03) & 0.25 & 0.28 (0.06)   \\
  & 3 & 0.25 & 0.23 (0.03) & 0.25 & 0.25  (0.03) & 0.10 & 0.15 (0.04)   \\  
  & 4 & 0.50 & 0.53 (0.04) & 0.15 & 0.17  (0.03) & 0.05 & 0.06 (0.03)   \\
8 & 1 & 0.15 & 0.20 (0.03) & 0.20 & 0.20  (0.03) & 0.55 & 0.67 (0.07)  \\
  & 2 & 0.20 & 0.14 (0.03) & 0.45 & 0.45  (0.04) & 0.20 & 0.16 (0.06)   \\
  & 3 & 0.20 & 0.20 (0.03) & 0.25 & 0.26  (0.03) & 0.15 & 0.12 (0.04)  \\  
  & 4 & 0.45 & 0.46 (0.04) & 0.10 & 0.10  (0.02) & 0.10 & 0.05 (0.03)   \\
9 & 1 & 0.40 & 0.35 (0.04) & 0.40 & 0.43 (0.04) & 0.40 &  0.39 (0.07)   \\
  & 2 & 0.50 & 0.53 (0.04) & 0.50 & 0.46  (0.04) & 0.50 & 0.47 (0.07)   \\
  & 3 & 0.10 & 0.12 (0.02) & 0.10 & 0.11  (0.02) & 0.10 & 0.14 (0.04)   \\
10 & 1 & 0.70 & 0.72 (0.04) & 0.70 & 0.72  (0.04) & 0.70 & 0.68 (0.08)   \\
  & 2 & 0.10 & 0.10 (0.02) & 0.10 & 0.11 (0.02) & 0.10 & 0.15 (0.04)  \\
  & 3 & 0.20 & 0.18 (0.03) & 0.20 & 0.17 (0.03) & 0.20 & 0.17 (0.05)  \\
11 & 1 & 0.20 & 0.21 (0.03) & 0.20 & 0.20 (0.03) & 0.20 & 0.20 (0.05)  \\
  & 2 & 0.20 & 0.18 (0.03) & 0.20 & 0.20 (0.03) & 0.20 & 0.15 (0.04)  \\
  & 3 & 0.20 & 0.22 (0.03) & 0.20 & 0.20 (0.03) & 0.20 & 0.23 (0.05)  \\
  & 4 & 0.20 & 0.21 (0.03) & 0.20 & 0.20 (0.03) & 0.20 & 0.20 (0.05)   \\
  & 5 & 0.20 & 0.19 (0.03) & 0.20 & 0.20 (0.03) & 0.20 & 0.22 (0.05)  \\
12 & 1 & 0.10 & 0.15 (0.03) & 0.10 & 0.11 (0.02) & 0.10 & 0.08 (0.03) \\
  & 2 & 0.15 & 0.17 (0.03) & 0.15 & 0.13 (0.02) & 0.15 & 0.16 (0.04)  \\
  & 3 & 0.20 & 0.16 (0.03) & 0.20 & 0.20 (0.03) & 0.20 & 0.25 (0.05)  \\
  & 4 & 0.25 & 0.24 (0.03) & 0.25 & 0.21 (0.03) & 0.25  & 0.21 (0.05)  \\
  & 5 & 0.30 & 0.28 (0.03) & 0.30 & 0.34 (0.03) & 0.30 & 0.31 (0.06)  \\
13 & 1 & 0.20 & 0.20 (0.03) & 0.20 & 0.20 (0.03) & 0.20 & 0.23 (0.05)  \\
  & 2 & 0.30 & 0.27 (0.03) & 0.30 & 0.27 (0.03) & 0.30 & 0.33 (0.06)   \\
  & 3 & 0.30 & 0.31 (0.04) & 0.30 & 0.30 (0.03) & 0.30 & 0.29 (0.06)  \\
  & 4 & 0.10 & 0.10 (0.02) & 0.10 & 0.14 (0.03) & 0.10 & 0.09 (0.03)  \\
  & 5 & 0.10 & 0.11 (0.02) & 0.10 & 0.09 (0.02) & 0.10 & 0.06 (0.03)  \\
\bottomrule
\end{tabular}
\caption{True and estimated values of the item probability parameters for the item selection model for dataset 2.}
  \label{table:sim2_item_prob_estimates_itemsel}
\end{table}

\begin{table}[H]
\centering
\begin{tabular}{cccccccccccccc}
\hline
Variable & 1 & 2 & 3 & 4 & 5 & 6 & 7 & 8 & 9 & 10 & 11 & 12 & 13 \\
\hline
PIP (item) & 1.00 & 1.00 & 1.00 & 1.00 & 1.00 & 1.00 & 1.00 & 1.00 & 0.01 & 0.00 & 0.00 & 0.00 & 0.00 \\
PIP (simul) & 1.00 & 1.00 & 1.00 & 1.00 & 1.00 & 1.00 & 1.00 & 1.00 & 0.01 & 0.00 & 0.00 & 0.00 & 0.00 \\
\hline
\end{tabular}
\caption{Posterior inclusion probabilities (PIP) for each item variable in the LCR model for dataset 2. 
PIP (item): item selection alone; PIP (simul): simultaneous selection.}\label{table:item_selection_results_sim_study2}
\end{table}

\section{CSHQ Sleep Data Analysis}
\subsection{Study Sample Characteristics}
Table \ref{table:demographic_diagnostic_table_CSHQ} contains demographic and diagnostic information for the sleep patterns dataset.
\begin{table}[H]
\centering
\begin{tabular}{llcc}
\toprule
 Demographic & &  $n$ & \% \\
\midrule
 Gender&  & \\
& Female &  69 & 46.6\\
& Male &   79 & 53.4\\
Diagnosis & & \\
& Neurotypical & 104 & 70.3 \\
& Neurodiverse & 44 & 29.7 \\
\bottomrule
Diagnosis/comorbidity (ND) & & $n$ & \% (ND group) \\
\midrule
ASD & & 19 & 43.18 \\
ID/ASD & & 6 & 13.64 \\
ADHD/ASD & & 5 & 11.36 \\
Dyslexia & & 3 & 6.82 \\
ADHD & & 2 & 4.55 \\
ASD/ADHD/Dyslexia & & 1 & 2.27 \\
ID/ASD/ADHD & & 1 & 2.27 \\
ASD/GDD & & 1 & 2.27 \\
ADHD/Dyslexia & & 1 & 2.27 \\
ID/Dyslexia & & 1 & 2.27 \\
ID & & 1 & 2.27 \\
Dyspraxia & & 1 & 2.27 \\
Tourette's & & 1 & 2.27 \\
Brain Abnormality & & 1 & 2.27 \\
\bottomrule
\end{tabular}
\caption{Demographic and diagnostic summary information for the CSHQ dataset (ND = neurodiverse, ASD = Autism Spectrum Disorder, ID = Intellectual Disability, GDD = Global Developmental Delay)}\label{table:demographic_diagnostic_table_CSHQ}
\end{table}
\subsection{Model Parameter Estimates}
Table \ref{table:CSHQ_item_prob_estimates_itemsel} contains estimates for the item probability parameters computed post-hoc for the sleep data LCR model with item selection. Table \ref{table:CSHQ_beta_estimates} contains posterior mean values for LCR regression coefficient parameters, along with sd and 95\% HDI values. Table \ref{table:CSHQ_beta_estimates_varsel} contains the estimates for the regression coefficient parameters for the ASD diagnosis variable with predictor selection.
\begin{longtable}{clccccc}
\caption{Estimated values of the item probability parameters for the item selection model for the CSHQ sleep data}
\label{table:CSHQ_item_prob_estimates_itemsel}\\
\toprule
\textbf{Variable} & & \textbf{Category}  & \textbf{Est. 1 (sd)}  & \textbf{Est. 2 (sd)}  & \textbf{Est. 3 (sd)} & \textbf{Est. 4 (sd)}  \\
\midrule
\endfirsthead

\multicolumn{7}{c}{\tablename\ \thetable\ -- \textit{Continued from previous page}} \\
\toprule
\textbf{Variable} & & \textbf{Category}  & \textbf{Est. 1 (sd)}  & \textbf{Est. 2 (sd)}  & \textbf{Est. 3 (sd)} & \textbf{Est. 4 (sd)}  \\
\midrule
\endhead

\midrule
\multicolumn{7}{r}{\textit{Continued on next page}} \\
\endfoot

\bottomrule
\endlastfoot
Bedtime Resistance &1 & 1  & 0.78 (0.06) & 0.77  (0.09) & 0.63 (0.10) & 0.68 (0.11)  \\
 & & 2  & 0.18 (0.05) & 0.18  (0.07) &  0.30 (0.09) & 0.27 (0.09)  \\
 & & 3  & 0.04 (0.03) &  0.05 (0.04) &  0.07 (0.05) & 0.04 (0.04) \\
&2 & 1  & 0.94 (0.03) &  0.75 (0.07) &  0.90 (0.05) & 0.51 (0.11)\\
 & & 2  & 0.02 (0.02) &  0.09 (0.05) &  0.07 (0.04) & 0.33 (0.10)\\
 & & 3  & 0.05 (0.03) &  0.16 (0.06) &  0.03 (0.03) & 0.16 (0.07)\\
&3 & 1  & 0.94 (0.03) &  0.75 (0.07) &  0.87 (0.06) & 0.43 (0.11)\\
&  & 2  & 0.02 (0.02) &  0.07 (0.04) &  0.09 (0.05) & 0.38 (0.10)\\
&  & 3  & 0.05 (0.03) &  0.18 (0.07) &  0.03 (0.03) & 0.19 (0.08)\\
 &5 & 1  & 0.89 (0.04) &  0.14 (0.07) &  0.77 (0.08) & 0.06 (0.04)\\
 & & 2  & 0.09 (0.04) &  0.38 (0.09) &  0.20 (0.08) & 0.25 (0.09)\\
&  & 3  & 0.02 (0.02) &  0.47 (0.08) &  0.04 (0.04) & 0.69 (0.10)\\

Sleep Onset Delay & & 1 & 0.58 (0.06) &  0.48  (0.09) & 0.15  (0.07) & 0.20 (0.09)\\
 & & 2 & 0.27 (0.06) &  0.41  (0.09) & 0.57  (0.09) & 0.35 (0.10)\\
 & & 3 & 0.15 (0.05) &  0.11  (0.05) &  0.28 (0.08) & 0.44 (0.10)\\
 
Sleep Duration  & 1 & 1  & 0.85 (0.05) &  0.65  (0.09) &  0.14 (0.07) & 0.08 (0.06)\\
 & & 2  & 0.12 (0.05) &  0.29  (0.09) &  0.55 (0.09) & 0.57 (0.11)\\
 & & 3  & 0.03 (0.02) &  0.06  (0.04) &  0.32 (0.09) & 0.35 (0.10)\\
  & 2 & 1  & 0.86 (0.05) &  0.65  (0.09) &  0.15 (0.07) & 0.05 (0.04)\\
 & & 2  & 0.10 (0.04) &  0.27  (0.09) & 0.53  (0.09) & 0.61 (0.10)\\
 & & 3  & 0.05 (0.03) &  0.09  (0.05) &  0.32 (0.09) & 0.35 (0.10)\\
   & 3 & 1  & 0.86 (0.05) &  0.74  (0.08) &  0.22 (0.08) & 0.29 (0.09)\\
 & & 2  & 0.11 (0.04) &  0.23  (0.08) &  0.54 (0.09) & 0.37 (0.10)\\
 & & 3  & 0.03 (0.02) &  0.03  (0.03) & 0.24  (0.08) & 0.33 (0.10)\\
 
 Sleep Anxiety  & 1 & 1  & 0.80 (0.05) &  0.14  (0.07) & 0.89  (0.06) & 0.09 (0.06) \\
 & & 2  & 0.13 (0.04) &  0.22  (0.07) &  0.06 (0.05) & 0.27 (0.05) \\
 & & 3  & 0.06 (0.03) &   0.63 (0.09) &  0.05 (0.04) & 0.64 (0.04) \\
  & 2 & 1  & 0.70 (0.06) &  0.35  (0.08) & 0.67  (0.09) & 0.31 (0.09) \\
 & & 2  & 0.20 (0.05) &  0.19  (0.07) &  0.12 (0.06) & 0.36 (0.10)\\
 & & 3  & 0.10 (0.04) &  0.46  (0.09) &  0.21 (0.08) & 0.32 (0.10)\\
   & 3 & 1  & 0.89 (0.04) &   0.11 (0.06) &  0.76 (0.08) & 0.06 (0.04)\\
 & & 2  & 0.08 (0.04) &  0.35  (0.08) &  0.20 (0.08) & 0.35 (0.10)\\
 & & 3  & 0.02 (0.02) &  0.54  (0.09) &  0.04 (0.04) & 0.59 (0.10)\\
    & 4 & 1  & 0.70 (0.06) &  0.63  (0.08) &  0.41  (0.09) & 0.14 (0.08) \\
 & & 2  & 0.27 (0.06) &  0.26  (0.07) & 0.40  (0.09) & 0.47 (0.10)  \\
 & & 3  & 0.03 (0.02) &  0.11  (0.05) & 0.19  (0.07) & 0.39 (0.10) \\
 
Night Waking  & 1 & 1  & 0.90 (0.04) &  0.48  (0.09) & 0.71  (0.09) & 0.18 (0.08)\\
 & & 2  & 0.08 (0.04) &  0.34  (0.08) &  0.20 (0.08) & 0.34 (0.10)\\
 & & 3  & 0.02 (0.02) &  0.17  (0.07) &  0.10 (0.07) & 0.48 (0.10)\\
  & 2 & 1  & 0.76 (0.06) &  0.33  (0.08) & 0.31  (0.08) & 0.10 (0.06)\\
 & & 2  & 0.21 (0.06) &  0.46  (0.09) &  0.33 (0.09) & 0.27 (0.09)\\
 & & 3  & 0.03 (0.02) &  0.21  (0.07) &  0.35 (0.09) & 0.63 (0.10)\\
   & 3 & 1  & 0.94 (0.03) &  0.64  (0.08) & 0.41  (0.10) & 0.36 (0.10)\\
 & & 2  & 0.04 (0.03) &  0.33  (0.08) &  0.25 (0.08) & 0.39 (0.10)\\
 & & 3  & 0.02 (0.02) &  0.03  (0.03) & 0.33  (0.09) & 0.25 (0.09)\\
 
 Parasomnias  & 1 & 1  & 0.91  (0.06) &  0.87  (0.09) &  0.81 (0.10) & 0.72 (0.11) \\
 & & 2  & 0.05 (0.03) &  0.08  (0.04) & 0.09  (0.05) & 0.12 (0.07)\\
 & & 3  & 0.05 (0.03) &  0.05 (0.04) &  0.09 (0.05) & 0.16 (0.07)\\
  & 2 & 1  & 0.68  (0.06) &  0.51  (0.09) & 0.77  (0.09) & 0.43 (0.10)\\
 & & 2  & 0.26 (0.06) &  0.46  (0.09) & 0.13  (0.07) & 0.49 (0.11)\\
 & & 3  & 0.05 (0.03) &  0.03  (0.03) & 0.10  (0.06) & 0.08 (0.06)\\
   & 3 & 1  & 0.48 (0.06) &  0.28  (0.08) & 0.27  (0.08) & 0.16 (0.08)\\
 & & 2  & 0.43 (0.06) &  0.54  (0.08) & 0.35  (0.09) & 0.34 (0.10)\\
 & & 3  & 0.10 (0.04) &  0.19  (0.07) & 0.38  (0.09) & 0.50 (0.10)\\
    & 4 & 1  & 0.95  (0.05) &  0.91  (0.09) & 0.88  (0.10) & 0.83 (0.10)\\
 & & 2  & 0.05 (0.03) &  0.09  (0.05) & 0.12  (0.06) & 0.17 (0.08)\\
    & 5 & 1  & 0.89 (0.04) &  0.67  (0.09) & 0.75  (0.08) & 0.29 (0.09)\\
 & & 2  & 0.06 (0.03) &  0.18  (0.07) & 0.16  (0.07) & 0.41 (0.10)\\
 & & 3  & 0.06 (0.03) &  0.16  (0.06) & 0.09  (0.06) & 0.30 (0.09)\\
    & 6 & 1  & 0.93  (0.05) &  0.90  (0.08) & 0.79  (0.10) & 0.73 (0.11)\\
 & & 2  & 0.07 (0.03) &  0.10  (0.05) & 0.21  (0.08) & 0.27 (0.09)\\
    & 7 & 1  & 0.80 (0.06) &  0.69  (0.09) & 0.68  (0.10) & 0.60 (0.11)\\
 & & 2  & 0.19 (0.05) &  0.26  (0.08) & 0.26  (0.08) & 0.35 (0.10)\\
 & & 3  & 0.02 (0.02) &  0.05 (0.04) &  0.06 (0.04) & 0.04 (0.04)\\
 Sleep Disordered Breathing &1 & 1  & 0.82 (0.06) & 0.72  (0.10) &  0.72 (0.10) & 0.62 (0.11) \\
 & & 2  & 0.14 (0.04) &  0.22 (0.07) & 0.16  (0.07) & 0.34 (0.10) \\
 & & 3  & 0.05 (0.03) &  0.05 (0.04) & 0.12  (0.06) & 0.04 (0.04) \\
&2 & 1  & 0.98 (0.05) &  0.97 (0.08) & 0.90  (0.09) & 0.83 (0.10) \\
 & & 2  & 0.02 (0.02) &  0.03 (0.03) & 0.10  (0.05) & 0.17 (0.08) \\
&3 & 1  & 0.92 (0.05) &  0.94 (0.08) & 0.79  (0.10) & 0.80 (0.11) \\
&  & 2  & 0.08 (0.04) &  0.06 (0.04) & 0.21  (0.08) & 0.20 (0.08) \\

Daytime Sleepiness &1 & 1  & 0.45 (0.07) & 0.66  (0.09) & 0.46  (0.10) & 0.53 (0.11) \\
 & & 2  & 0.35 (0.06) & 0.16  (0.07) & 0.43  (0.10) & 0.30 (0.10) \\
 & & 3  & 0.20 (0.05) & 0.18  (0.07) & 0.11  (0.06) & 0.17 (0.08) \\
&2 & 1  & 0.68 (0.06) & 0.77  (0.08) & 0.16  (0.07) & 0.22 (0.09) \\
 & & 2  & 0.28 (0.06) & 0.20  (0.08) & 0.63  (0.09) & 0.53 (0.10) \\
 & & 3  & 0.03 (0.02) & 0.03  (0.03) & 0.21  (0.07) & 0.25 (0.09) \\
&3 & 1  & 0.59 (0.07) & 0.75  (0.09) & 0.30  (0.09) & 0.34 (0.10) \\
&  & 2  & 0.33 (0.06) & 0.22  (0.08) & 0.34  (0.09) & 0.44 (0.10) \\
&  & 3  & 0.08 (0.04) & 0.03  (0.03) & 0.36  (0.07) & 0.21 (0.08) \\
&4 & 1  & 0.59 (0.06) & 0.75  (0.08) & 0.30  (0.08) & 0.34 (0.10) \\
&  & 2  & 0.37 (0.06) & 0.22  (0.08) & 0.34  (0.09) & 0.44 (0.10) \\
&  & 3  & 0.03 (0.02) & 0.03  (0.03) & 0.36  (0.09) & 0.21 (0.08) \\
&5 & 1  & 0.84 (0.05) & 0.81  (0.07) & 0.42  (0.09) & 0.33 (0.10) \\
&  & 2  & 0.14 (0.05) & 0.16  (0.06) & 0.46  (0.09) & 0.55 (0.11) \\
&  & 3  & 0.02 (0.02) & 0.03  (0.03) & 0.12  (0.06) & 0.12 (0.07) \\
&6 & 1  & 0.62  (0.06) & 0.60  (0.09) & 0.16  (0.07) & 0.09 (0.06) \\
&  & 2  & 0.36 (0.06) & 0.38  (0.09) & 0.51  (0.10) & 0.59 (0.10) \\
&  & 3  & 0.02 (0.02) & 0.03  (0.03) & 0.33  (0.09) & 0.31 (0.10) \\
&7 & 1  & 0.92 (0.06) & 0.90  (0.09) & 0.78  (0.10) & 0.80 (0.11) \\
&  & 2  & 0.05 (0.03) & 0.05  (0.04) & 0.12  (0.06) & 0.04 (0.04) \\
&  & 3  & 0.03 (0.02) & 0.05  (0.04) & 0.09  (0.05) & 0.16 (0.07) \\
&8 & 1  & 0.82 (0.06) & 0.77  (0.09) & 0.76  (0.10) & 0.53 (0.11) \\
&  & 2  & 0.17 (0.05) & 0.17  (0.07) & 0.21  (0.08) & 0.36 (0.10) \\
&  & 3  & 0.02 (0.02) & 0.05  (0.04) & 0.03 (0.03) &  0.12 (0.06)\\
\bottomrule

\end{longtable}

\begin{table}[H]
\centering
\begin{tabular}{clrcc}
\hline
\textbf{Group} & \textbf{Parameter} & \textbf{Post. Mean} & \textbf{Post. SD} & \textbf{95\% HDI}  \\
\hline
\multirow{7}{*}{B} & $\bsbet_{\text{B}0}$ (intercept) & $-0.32$  & $0.37$ & $[-1.08, 0.37]$   \\
&$\bsbet_{\text{B}1}$ (age) & $-0.65$ & $0.37$ & $[-1.38, -0.05]$     \\
&$\bsbet_{\text{B}2}$ (gender) & $-0.93$  & $0.67$ & $[-2.33, 0.34]$      \\
&$\bsbet_{\text{B}3}$ (ASD diagnosis) & $1.88$ & $1.08$ &  $[-0.09, 4.10]$     \\
&$\bsbet_{\text{B}4}$ (ID diagnosis) & $-2.72$ & $3.65$ &  $[-10.38, 3.62] $    \\
&$\bsbet_{\text{B}5}$ (other diagnoses) & $-5.39$ & $3.23$ &  $[-11.97, -0.19]$     \\
\hline
\multirow{7}{*}{C} &$\bsbet_{\text{C}0}$ (intercept) &  $-1.45$ & $0.44$ &  $[-2.29, -0.58]$  \\
&$\bsbet_{\text{C}1}$ (age) & $0.24$ & $0.33$ &  $[-0.39, 0.91]$    \\
&$\bsbet_{\text{C}2}$ (gender) & $0.02$ & $0.70$ & $[-1.32, 1.38]$     \\
&$\bsbet_{\text{C}3}$ (ASD diagnosis) & $2.88$ & $0.93$ &  $[1.07, 4.71]$     \\
&$\bsbet_{\text{C}4}$ (ID diagnosis) & $2.27$ & $1.75$ &  $[-0.92, 5.75]$    \\
&$\bsbet_{\text{C}5}$ (other diagnoses) & $-0.55$ & $1.07$ &  $[-2.52, 1.61]$    \\
\hline
\multirow{7}{*}{D} &$\bsbet_{\text{P}0}$ (intercept) & $-1.60$ & $0.33$ & $[-2.26, -0.94] $  \\
&$\bsbet_{\text{D}1}$ (age) & $-0.12$ & $0.27$ & $[-0.65, 0.41]$     \\
&$\bsbet_{\text{D}2}$ (gender) & $0.02$ & $0.60$ &  $[-1.16, 1.19]$    \\
&$\bsbet_{\text{D}3}$ (ASD diagnosis) & $3.40$ & $0.85$ & $[1.81, 5.09]$     \\
&$\bsbet_{\text{D}4}$ (ID diagnosis) & $0.53$ & $1.79$ &   $[-2.75, 4.20]$   \\
&$\bsbet_{\text{D}5}$ (other diagnoses) & $-2.05$ & $0.94$ &   $[-2.90, 0.77]$    \\
\hline
\end{tabular}
\caption{Summary table containing the posterior mean estimated from MCMC samples for the CSHQ data using only item selection, along with the 95\% HDI and posterior SD. The gender variable was binary encoded, taking value 1 for a response of female.}\label{table:CSHQ_beta_estimates}
\end{table}

\begin{table}[H]
\centering
\begin{tabular}{clrcc}
\hline
\textbf{Group} & \textbf{Parameter} & \textbf{Post. Mean} & \textbf{Post. SD} & \textbf{95\% HDI}  \\
\hline
\multirow{2}{*}{B} & $\bsbet_{\text{B}0}$ (intercept) & $-0.60$  & $0.24$ & $[-1.06,-0.16]$   \\
&$\bsbet_{\text{B}3}$ (ASD diagnosis) & $0.82$ & $0.89$ &  $[-0.98, 2.51]$     \\
\hline
\multirow{2}{*}{C} &$\bsbet_{\text{C}0}$ (intercept) &  $-1.33$ & $0.35$ &  $[-2.01, -0.63]$  \\
&$\bsbet_{\text{C}3}$ (ASD diagnosis) & $2.96$ & $0.89$ &  $[1.22, 4.69]$     \\
\hline
\multirow{2}{*}{D} &$\bsbet_{\text{D}0}$ (intercept) & $-1.67$ & $0.37$ & $[-2.42, -0.97]$   \\
&$\bsbet_{\text{D}3}$ (ASD diagnosis) & $2.30$ & $0.87$ & $[1.39, 4.82] $    \\
\hline
\end{tabular}
\caption{Summary table containing the posterior mean estimated from MCMC samples for the CSHQ data using full variable selection, along with the 95\% HDI and posterior SD. Estimates are computed over samples taken from iterations when the ASD diagnosis variable was included in the model.}\label{table:CSHQ_beta_estimates_varsel}
\end{table}

\subsection{Variable Selection Results}
Table \ref{table:item_selection_results_CSHQ} contains posterior inclusion probabilities for each item variable in the CSHQ survey.

\begin{table}[H]
\centering
\begin{tabular}{ccccc}
\hline
CSHQ Subscale & Variable & PIP (LCR - Item) & PIP (LCR - simul.) & PIP (LCA) \\
\hline
Bedtime Resistance & 1 & 0.00 & 0.00 & 0.00   \\
&2 &  0.96 & 0.99 &  1.00  \\
&3 &  1.00 & 1.00 &  1.00  \\
&5 &  1.00 & 1.00 &  1.00  \\
\hline
Sleep Onset Delay & & 1.00 & 1.00 &  1.00    \\
\hline
Sleep Duration &1 & 1.00 & 1.00 &  1.00  \\
& 2 & 1.00 & 1.00 & 1.00 \\
&3 & 1.00 & 1.00 & 1.00  \\
\hline
Sleep Anxiety&1 & 1.00 & 1.00 & 1.00   \\
&2 & 0.99 & 1.00 &  1.00 \\
&3 & 1.00 & 1.00 & 1.00  \\
&4 & 1.00 & 1.00 & 1.00  \\
\hline
Night Waking&1 & 1.00 & 1.00 & 1.00   \\
&2 & 1.00 & 1.00 &  1.00  \\
&3 & 1.00 & 1.00 &  1.00  \\
\hline
Parasomnias&1 & 0.00 & 0.00 &  0.00  \\
&2 & 0.50 & 0.72 &  0.91  \\
&3 & 0.99 & 1.00 & 1.00   \\
&4 & 0.02 & 0.02 &  0.00  \\
&5 & 1.00 & 1.00 &  1.00  \\
&6 & 0.28 & 0.27 &  0.45  \\
&7 & 0.00 & 0.00 &  0.00  \\
\hline
Sleep Disordered Breathing&1 & 0.01 & 0.00 &  0.00  \\
&2 & 0.08 & 0.09 & 0.01  \\
&3 & 0.13 & 0.08 &  0.01  \\
\hline
Daytime Sleepiness&1 & 0.15 & 0.19 & 0.12   \\
&2 & 1.00 & 1.00 &  1.00  \\
&3 & 0.48 & 0.42 &  0.29  \\
&4 & 1.00 & 1.00 &  1.00  \\
&5 & 1.00 & 1.00 &  1.00  \\
&6 & 1.00 & 1.00 &  1.00  \\
&7 & 0.00 & 0.00 &  0.00  \\
&8 & 0.01 & 0.00 &  0.00  \\
\hline
\end{tabular}
\caption{Table containing the posterior inclusion probabilities (PIP) for each of the CSHQ items, from the item selection model for latent class regression (LCR - Item), the simultaneous item variable and predictor variable selection (LCR - simul.) and from the collapsed latent class analysis model (LCA)}\label{table:item_selection_results_CSHQ}
\end{table}

\subsection{Supplementary Visualisations}
\begin{figure}[H]
    \centering
    \includegraphics[width=\linewidth]{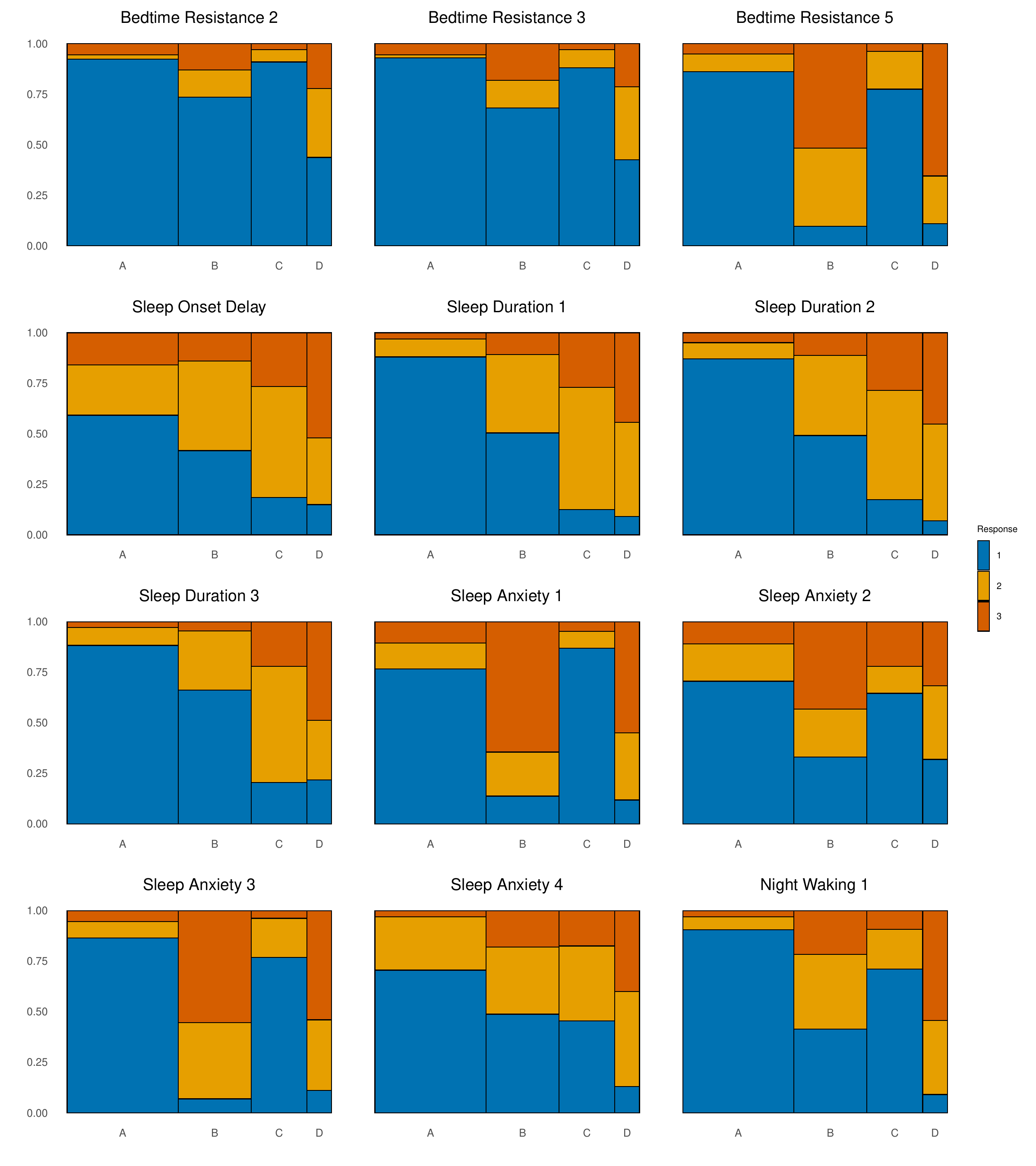}
    \caption{}\label{fig:CSHQ_full_mosaic1}
\end{figure}

\begin{figure}[H]
    \centering
    \includegraphics[width=\linewidth]{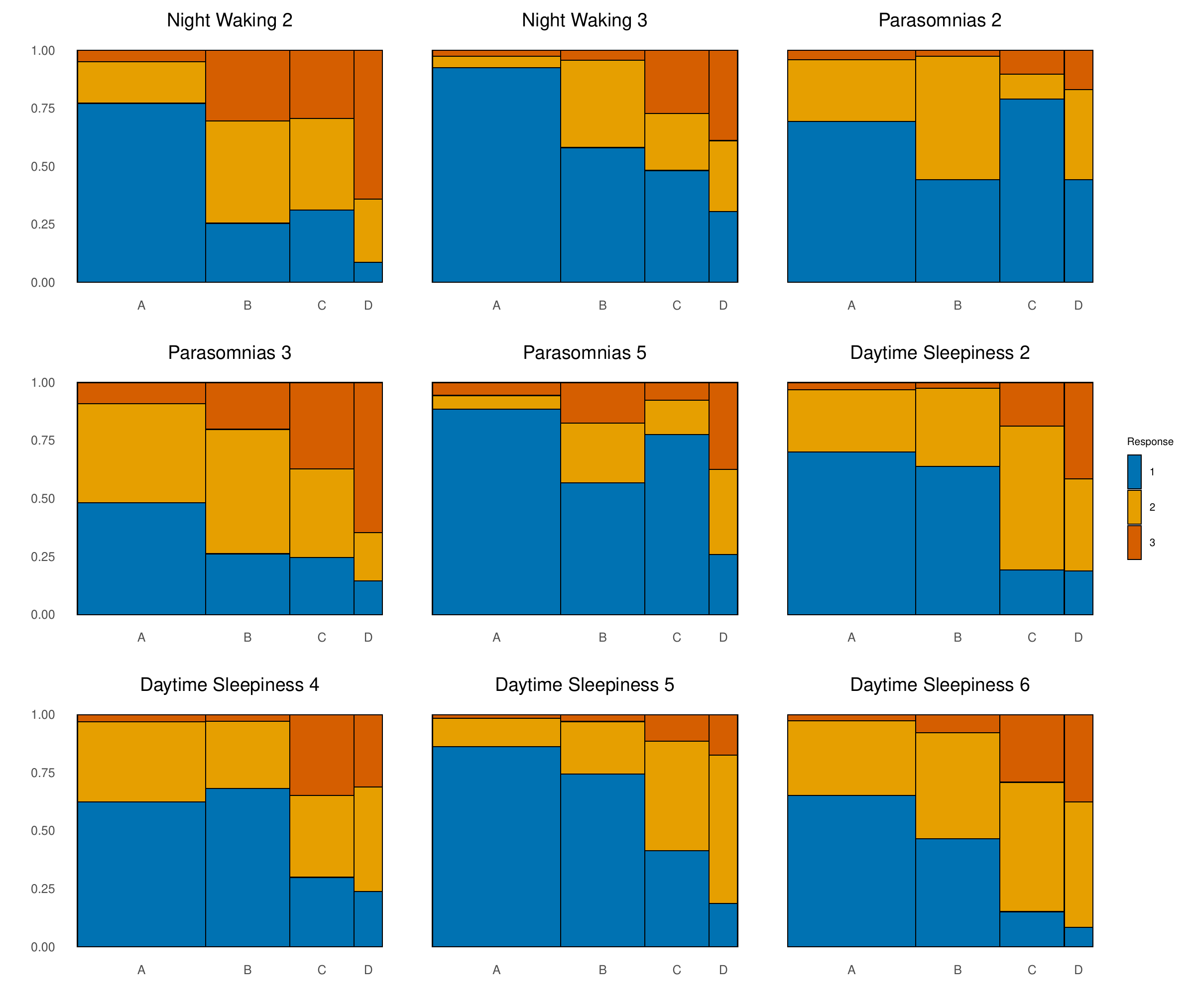}
    \caption{}\label{fig:CSHQ_full_mosaic2}
\end{figure}

Figures \ref{fig:CSHQ_full_mosaic1} and \ref{fig:CSHQ_full_mosaic2} contain mosaic plots for the each of the item variables of the CSHQ sleep patterns dataset which were retained by the collapsed sampler based on the posterior inclusion probabilities being greater than 0.5.

\section{Technical Derivations}

\subsection{Derivation of Pólya-Gamma Sampling Method for LCR}
This follows the method of \cite{polson_bayesian_2013}. Consider the full conditional for $\bsbet$ from (4). We can consider each of the $\bsbet_g$ parameters conditional on $\bsbet_{-g}$, the $\bsbet$ matrix with the $g$-th column removed. We write the expression for the full conditional distribution of $\bsbet_g$ as
\begin{equation}\label{beta_fc2}
p(\bsbet_g \mid \bsbet_{-g}, \boldX, \boldY) \propto p(\bsbet_g) \prod_{i=1}^n \left[  \frac{\exp(\boldX_i^T \bsbet_g)}{\sum_{h=1}^G \exp(\boldX_i^T \bsbet_h)}\right]^{z_{ig}}\left[1- \frac{\exp(\boldX_i^T \bsbet_g)}{\sum_{h=1}^G \exp(\boldX_i^T \bsbet_h)}\right]^{1-z_{ig}}.
\end{equation}
This can be rewritten in the form
\[
p(\bsbet_g \mid \bsbet_{-g}, \boldX, \boldZ) \propto p(\bsbet_g) \prod_{i=1}^n \frac{\exp(\eta_{ig})^{z_{ig}}}{1+\exp(\eta_{ig})},
\]
where
\[
\eta_{ig} = \boldX_i^T\bsbet_g - C_{ig}, \qquad C_{ig} = \log\left( \sum_{h \neq g} \exp(\boldX_i^T \bsbet_h) \right).
\]
By utilising the result from (5), we can introduce the Pólya-Gamma random variables $\omega_{ig} \sim \pg(1,0)$ for each $i = 1,\dots,n, g = 1,\dots,G-1$, which allows us to rewrite the full conditional as 
\[
p(\bsbet_g \mid \bsbet_{-g},\boldX, \boldY) 
\propto p(\bsbet_g) \prod_{i=1}^n \exp\left( \kappa_{ig}\eta_{ig} - \frac{\omega_{ig}\eta_{ig}^2}{2} \right),
\]
where $\kappa_{ig} = z_{ig} - \frac{1}{2}$. We can recognise the second multiplicative term here as a product of Gaussian kernels in the $\eta$ variables. By setting the prior $\bsbet_g \sim \cN(\mathbf{m}_{0g}, \mathbf{V}_{0g})$, we are left with a two part update for $\bsbet_g$ and $\omega_{ig}$, given by
 \begin{equation}
 \begin{aligned}
    \bsbet_g \mid \bsbet_{-g}, \omega_{\cdot g}, \boldX,\boldZ  &\sim \cN(\mathbf{m}_g, \mathbf{V}_g),\\
    \omega_{ig} \mid \bsbet, \boldX_i &\sim \pg(1,\eta_{ig}).
\end{aligned}
\end{equation}
where
\[
\mathbf{V}_g^{-1} = \boldX^T \Omega_g \boldX + \boldV_{0g}^{-1}, \qquad \boldm_g = \boldV_g \left( \boldX^T(\bskap_g+ \Omega_g\boldC_{\cdot g}) + \boldV_{0g}^{-1}\boldm_{0g} \right), \qquad \Omega_g = \diag(\{\omega_{ig}\}_{i=1}^n),
\]
and $\boldC_{\cdot g}$ is the $g$-th column of $\boldC = (C_{ig})$.

\bibliographystyle{apalike-doi} 
\bibliography{references}